\documentclass[11pt, oneside]{article}   	
\usepackage{geometry}                		
\usepackage{graphicx}				

\usepackage{hyperref}
\usepackage{amssymb}
\usepackage{tikz}
\usepackage{latexsym, amsmath, mathrsfs, graphicx}
\usepackage[utf8]{inputenc}
\newcommand{\beq}{\begin{eqnarray}}
\newcommand{\eeq}{\end{eqnarray}}
\usepackage{authblk}
\usepackage{xcolor}
\usepackage{esint}
\usepackage{empheq}
\usepackage{subcaption}
\usepackage[normalem]{ulem}

\title{Thermodynamic bootstrap program for integrable QFT's: \\ Form factors and correlation functions at finite energy density}
\author{Axel Cort\'{e}s Cubero
\footnote{a.cortescubero@uu.nl}}
\affil{Institute for Theoretical Physics, Center for Extreme Matter and Emergent Phenomena,
Utrecht University, Princetonplein 5, 3584 CC Utrecht, the Netherlands}
\author{Mi\l{}osz Panfil \footnote{milosz.panfil@fuw.edu.pl}}
\affil{ Faculty of Physics, University of Warsaw, ul. Pasteura 5, 02-093 Warsaw, Poland}
\date{}							

\begin{document}
\maketitle

\begin{abstract}
We study the form factors of local operators of integrable QFT's between states with finite energy density. These states arise, for example, at finite temperature, or from a generalized Gibbs ensemble. We generalize Smirnov's form factor axioms, formulating them for a set of particle/hole excitations on top of the thermodynamic background, instead of the vacuum. We show that exact form factors can be found as minimal solutions of these new axioms. The thermodynamic form factors can be used to construct correlation functions on thermodynamic states. The expression found for the two-point function is similar to the conjectured LeClair-Mussardo formula, but using the new form factors dressed by the thermodynamic background, and with all singularities properly regularized. We study the different infrared asymptotics of the thermal two-point function, and show there generally exist two different regimes, manifesting massive exponential decay, or effectively gapless behavior at long distances, respectively. As an example, we compute the few-excitations form factors of vertex operators for the sinh-Gordon model.
\end{abstract}

\section{Introduction}

 One of the ultimate goals in the study of quantum many-body systems and quantum field theory (QFT) is the computation of correlation functions of local (physical) operators. Correlation functions are directly related to observable quantities such as scattering amplitudes in high energy physics or response functions in condensed matter settings. Correlation functions are also the key quantity to understand the emergence of macroscopic behavior from the underlying quantum microscopic description. 

One long-standing challenge has been the determination of correlation functions in strongly correlated quantum field theories at finite energy densities. Such situations arise naturally in condensed matter settings, where  systems at finite temperataure or finite chemical potential (e.g. biased systems) are commonly studied~\cite{CMFT_BOOK}. Similar problems are also of interest in high energy physics and cosmology. Some prominent examples are the study of hot and dense quark matter inside neutron stars~\cite{2018RPPh...81e6902B}, and the process of quark-gluon plasma formation in heavy ion collisions~\cite{LQM_BOOK,Arnold:2007pg}.

Recent years have also seen a renewed interest in the non-equilibrium physics, especially in low dimensions, of   QFT's and other quantum many-body systems \cite{RevModPhys.83.863,1742-5468-2016-6-064001,1742-5468-2017-1-013103}. Driven by questions of thermalization in quantum physics, this line of research led to a wider realization that, at least in principle, it is feasible to realize many exotic states of matter. They are exotic because they are very far from the vacuum, and cannot be described by traditional equilibrium thermodynamics. Examples of such states are those characterized by the generalized Gibbs ensembles \cite{1742-5468-2016-6-064007} (the precise structure of the GGE in QFT's has been studied in \cite{PhysRevA.91.051602,1742-5468-2017-2-023105,1742-5468-2017-1-013103,1742-5468-2017-2-023101}), inhomogeneous quasi-stationary steady states \cite{PhysRevX.6.041065, PhysRevLett.117.207201}, or even states capturing parts of the dynamics (Floquet dynamics) in periodically driven systems \cite{PhysRevE.90.012110,SciPostPhys.2.3.021,Cubero:2018hqt}. 

The natural question then arises: how can we compute correlation functions in such far from the vacuum, highly-excited states, characterized by a finite energy density. Especially, that the standard QFT techniques were mostly developed to tackle computation of the vacuum correlation functions.

In this work we address this problem for a specific subset of QFT's: integrable QFT's in $1+1$ dimensions. To achieve the aim of computing the correlation functions we develop a generalization of the form factor bootstrap approach~\cite{doi:10.1142/1115}. In this approach computation of the correlation function is divided in two steps. In the first step one has to determine form factors - matrix elements of the operator under the consideration. The second step is the evaluation of the spectral sum whose main ingredients are the form factors. The main result of our work is the generalization of the form-factor bootstrap program for states with finite energy density. This bootstrap program allows us to determine the form factors and realize the first step of the outlined above procedure. We then present an expression for the two-point correlation functions as an expansion in terms of these new form factors. The actual explicit evaluation of leading terms of the correlation functions is still a demanding analytic problem, as the form factors have singularities that need to be handled with care, however, we present an unambiguous prescription for regularizing these singularities.

\subsection{Integrable QFT's, bootstrap program and correlation functions}

Integrable field theories (IQFT) in 2d form an interesting subset of QFT's. They are characterized by elastic and factorizable scattering processes. Elasticity means that the set of particles momenta in a given state is fully conserved after any scattering process, and factorizability means in this context that any multiparticle scattering process can be written as a product over 2-particles scattering matrices.
In turn, these features are related to the existance of an infinite family of local conserved charges. This rich structure allows often for an exact determiantion of the S-matrix \cite{ZAMOLODCHIKOV1979253}, just from a few physical principles. 

Besides the intrinsic mathematical interest in IQFT's as exactly solvable toy models, they have also found numerous applications in a wide variety of physical problems. To give a small and not exhaustive list of examples, IQFT's can be used to describe the low energy dynamics of different quantum spin chains \cite{Essler:2004ht} and systems of cold atomic gases \cite{PhysRevB.75.174511} which are relevant to condensed matter experiments. Integrable QFT's can also be viewed as specific perturbation of  2d conformal field theories \cite{ZAMOLODCHIKOV1989641}, and provide a nonperturbative approach to study near-critical systems.  Applications in high energy physics include a recent proposal, showing that the worldsheet dynamics of the confing string in quantum chromodynamics can be approximately described by a massless 2d IQFT \cite{Cooper2015}, as well as numerous applications of integrability in the study of planar $\mathcal{N}=4$ supersymmetric Yang-Mills theory and the Ads/CFT correspondence \cite{Beisert:2010jr}.

One of the cornerstones in the development of the IQFT was construction of the form-factor bootstrap program \cite{KAROWSKI1978455,doi:10.1142/1115}. Following from a few physical principles, the bootstrap program allows to significantly constrain the form factors of local operators between states involving a finite number of particls on top of the vaccuum. In many circumstances the constraints are so strong that they allow the complete determination of the form factors, without relying on a Lagrangian formulation, or perturbation theory. Once these form factors are known, it is easy to write expressions for correlation functions as a form factor expansion. As a result, the vacuum correlation functions of integrable QFT's are, at least in principle, exactly computable. In practice, for models with rich spectrum and complicated scattering determinantion of the vaccumm correlation function might still be technically difficult.

There have been several previous approaches to computing correlation functions of integrable QFT's in finite-energy-density states. In particular, in Ref. \cite{LECLAIR1999624}, Leclair and Mussardo conjectured a general expression for one-point functions of local operators at finite temperature, in terms of a sum of standard, zero-temperature form factors.  A similar expression in terms of zero-temperature form factors was then also conjectured in \cite{LECLAIR1999624}, attempting to compute the two-point function at finite temperature.

Later independent checks have proven more rigorously that the Leclair-Mussardo conjecture is indeed correct for computing one-point functions \cite{POZSGAY2008209} (though some of concepts required for this proof were discovered as far back as in \cite{Balog:1992gf}). The validity of the two-point function proposal, however, has been much more contested, with some counter examples first explored in \cite{Saleur:1999hq} and \cite{CastroAlvaredo:2002ud}. 

A more fundamental problem with the two-point Leclair-Mussardo proposal was first addressed in \cite{Pozsgay:2010cr}. Namely, the form factor expansion proposed in \cite{LECLAIR1999624} contains some  divergent contributions, arising from kinematical poles of the form factors, and no clear unambiguous proposal is given on how to regularize this expression. It has been suggested that the Leclair-Mussardo two-point formula is only valid for cases where the kinematic poles do not present a problem, and the thermal background has no effect of dressing the particles of the theory, the only known examples being free theories \cite{Leclair:1996bf, CastroAlvaredo:2002ud} and field theories in 'tHooft's planar large-$N$ limit \cite{Cubero:2015ima}.

While the Leclair-Mussardo proposal for the two-point function, which in principle is an expression applicable to all values of the temperature, does not seem to be valid, some attempts have been successful at developing low temperature expansions \cite{Essler:2009zz,Pozsgay:2010cr}. These approaches consist simply on expanding the Gibbs-ensemble expressions at low temperatures, and carefully handling the kinematical singularities order by order in this expansion. These expansions naturally are expressed in terms of standard zero-temperature form factors, since the low-temperature expansion is centered around the zero-temperature ground state. These approaches do not seem to provide any viable pathway towards computing  correlators at higher temperature values, except if one is somehow able to resum an infinite number of terms.

A new approach towards two-point functions in highly excited states was recently developed in \cite{Pozsgay2018}, by showing a new way to regularize the kinematical singularities of form factors at all orders in the low-temperature expansion. This approach consisted on treating the two operators in the two-point function as a single quasilocal operator, whose form factors have similar kinematical pole structure as those of local operators. One then can rely on the results of the one-point Leclair-Mussardo formula, which provide the prescription to regularize kinematical poles of the quasilocal operator. While this approach is conceptually groundbreaking, solving the puzzle of how the kinematic poles can be regularized to all orders, pragmatically, it can still only be used as a low-temperature expansion centered around the zero-temperature ground state, unless one can find another way to resum higher order terms.

We finally point out that there has been recent progress in the computation of thermodynamic correlation functions numerically through the truncated conformal spectrum approach (TCSA) \cite{Yurov:1989yu}. This approach consists on treating the interacting QFT as a perturbation of a conformal field theory, whose finite-volume spectrum, and matrix elements are known. At finite volume, the spectrum becomes discrete, an can be truncated at some high-energy cutoff. The interacting Hamiltonian is then a finite, discrete matrix that can be diagonalized numerically. Only recently \cite{Kukuljan:2018whw}, the TCSA has been applied towards the computation of correlation functions in highly excited states, for the sine-Gordon model. We will concentrate only on simpler QFT's with {\it diagonal} scattering, such that the  computation of sine-Gordon correlators is beyond the scope of this paper, so we will not be able, at present, to make a quantitative comparison with this new numerical approach.

The approach we will develop in this paper has the novel feature that it is not a low-temperature expansion, but the expansion can be centered around any given finite energy density state.

\subsection{Summary of the results}

The starting point of our approach is the idea that thermal expectation values of {\it local} operators (also expectation values in generalized Gibbs ensembles) are equivalent in the thermodynamic limit (where the system size, $L$, goes to infinity) to the expectation value of the operator in just one single representative eigenstate. The representative eigenstate, which we denote as $\vert \vartheta\rangle$ is chosen such that the particle occupation number is the same when evaluated in the ensemble average, or in the single eigenstate,
\beq
\vartheta(\theta)\sim\lim_{L\to\infty}\langle A^\dag(\theta)A(\theta)\rangle_{\rm Thermal\,or\,GGE\,average}=\frac{\langle \vartheta\vert A^\dag(\theta)A(\theta)\vert \vartheta\rangle}{\langle\vartheta\vert\vartheta\rangle},\label{representativestate}
\eeq
where $A^\dag(\theta),\,A(\theta)$ are creation and annihilation operators, respectively, of a particle with rapidity $\theta$. The idea of the representative states orginated during the early studies of integrable models~\cite{korepin_bogoliubov_izergin_1993}. Since then it was applied  succesfully in numerous situations. These include studies of thermal correlation functions~\cite{Pozsgay:2010xd,2014_PRA_Panfil} and also studies of non-equilibrium QFT's, in an approach known as the ``quench action" method \cite{PhysRevLett.110.257203,2014_DeNardis_PRA_89,1742-5468-2014-10-P10035,Bertini:2016xgd}. To specify the representative state one then simply needs to specify the distribution $\vartheta(\theta)$.

Integrable field theories are characterized by a stable particle content (due to the property of elastic scattering). The multi-particle eigenstates in such theories are constructed simply by specifying rapidity $\theta$ of each particle. States of finite energy density are states in which there is an extensive number of particles such that the total energy of the state scales with the size of the system
\beq
  0 < \lim_{L\rightarrow\infty}\frac{1}{L} \frac{\langle \vartheta|H|\vartheta\rangle}{\langle\vartheta\vert\vartheta\rangle} < \infty,
\eeq
where $H$ is the QFT Hamiltonian. For simplicity of presentation, for the majority of this paper we assume that we are working with an IQFT with only one species of particle in the spectrum, though it is also possible to generalize our results for other models with a richer spectrum. To each state $|\vartheta\rangle$ we associate a smooth function $\vartheta(\theta)$ that plays a role of the filling function. The factor $\vartheta(\theta)d\theta$ determines how densely the interval $[\theta, \theta+d\theta]$ is filled with particles.

We can then consider a general two-point correlation function 
\begin{equation}
  C_{\vartheta}^{\mathcal{O},\mathcal{O}}(x,t) \equiv \frac{\langle\vartheta|\mathcal{O}(x,t)\mathcal{O}(0,0)|\vartheta\rangle}{\langle\vartheta\vert\vartheta\rangle},
\end{equation}
where for simplicity of the presentation we take both operators to be the same, which doesn't need to be the case in general. One way of handling the computation of such two point functions is by utilizing the Lehmann (spectral) representation. To this end, a crucial step is to assume that we can construct all the eigenstates $|\vartheta'\rangle$ such that
\begin{equation}
  \lim_{L\rightarrow\infty}\langle \vartheta'|\mathcal{O}(0)|\vartheta\rangle > 0.
\end{equation}
In other words we assume that we can span a subspace of the Hilbert space relevant for the computation of a given correlation function. Given that, we have
\begin{equation}
  C_{\vartheta}^{\mathcal{O},\mathcal{O}}(x,t) = \sum_{|\vartheta'\rangle \in \mathcal{H}} e^{it(E-E') - ix(P-P'))} |\langle\vartheta'|\mathcal{O}(0)|\vartheta\rangle|^2/\langle\vartheta\vert\vartheta\rangle.\label{lehmann}
\end{equation}
The matrix elements $\langle \vartheta'|\mathcal{O}(0)|\vartheta\rangle$, the form factors, are building blocks of the correlation function in the spectral representation. The aim of our work is to present an axiomatic way of computing such form factors in 2d integrable field theories, in the case where the states, $\vert \vartheta\rangle$ and $\vert \vartheta^\prime\rangle$ are states with finite energy density. 

The form factors $\langle \vartheta|\mathcal{O}(0)|\vartheta'\rangle$ are just matrix elements between two finite energy density states. For local physical operators, the form factor between two arbitrary states, $|\vartheta\rangle$ and $|\vartheta'\rangle$, is in most cases equal to zero because these two states will, in general, differ macroscopically: there will be an infinite number of particles that are present in one of the states and not in the other. Local physical operators cannot connect such macroscopically distant states. Therefore, the only kind of non-zero matrix element are those where the state $|\vartheta'\rangle$  is not too far from $|\vartheta\rangle$, and involves only microscopic deviations from this state. Therefore, $|\vartheta'\rangle$ must be given by the same filling function $\vartheta(\theta)$ plus some countable number of modifications: like adding or removing a particle with certain rapidity $\theta$. We write
\begin{equation}
  |\vartheta'\rangle \rightarrow |\vartheta, \theta_1, \dots, \theta_n\rangle.\label{varthetaprime}
\end{equation}
Note that the removal of a particle from the state $\vert \vartheta\rangle$ can be considered as the insertion of a ``hole" excitation. Given the relativistic invariance of IQFT's, a hole-excitation in the incoming state, is equivalent by crossing symmetry to a particle in the outgoing state. The crossing of a particle from incoming to outgoing states amounts to the transformation of its rapidity, $\theta\to\theta+\pi {\rm i}$, such that the notation (\ref{varthetaprime}) includes the insertion of both, particles and holes, as long as we allow $\pi {\rm i}$ shifts in rapidity. 

With this notation the form factors of local operators are defined as 
\begin{equation}
  f_{\vartheta}^{\mathcal{O}}(\theta_1, \dots, \theta_n) \equiv \frac{\langle\vartheta|\mathcal{O}(0)|\vartheta, \theta_1, \dots, \theta_n\rangle}{\langle\vartheta\vert\vartheta\rangle}.
\end{equation}
The filling function $\vartheta(\theta)$ plays a role of a background over which we construct additional excitations. We call such form factors ``finite background form factors". Sending the background to zero means considering form factors and excitations over the vacuum. 

To anticipate some of the results we will present now the most important axioms for the finite background form factors. Additional axioms are to be found in the text. The main axioms are:
\begin{itemize}
\item the scattering axiom
  \begin{equation}
   f_\vartheta^\mathcal{O}(\theta_1,\dots,\theta_i,\theta_{i+1},\dots,\theta_n)=S(\theta_i,\theta_i+1) f_\vartheta^\mathcal{O}(\theta_1,\dots,\theta_{i+1},\theta_i,\dots,\theta_n), 
  \end{equation}
\item the periodicity axiom
  \begin{equation}
    f_\vartheta^\mathcal{O}(\theta_1,\dots,\theta_n)=R_\vartheta(\theta_n\vert \theta_1,\dots,\theta_n) \,f_\vartheta^\mathcal{O}(\theta_n+2\pi{\rm i},\theta_1,\dots,\theta_{n-1}),
  \end{equation}
\item the annihilation-pole axiom
  \begin{align}
    -{\rm i}{\rm Res}_{\theta_1=\theta_2}f_\vartheta^\mathcal{O}(\theta_1+\pi {\rm i},\theta_2,\theta_3,\dots,\theta_n)
    =\left[1-{R}_\vartheta(\theta_2\vert\theta_3,\dots,\theta_n) \prod_{j=3}^n S(\theta_{2},\theta_j)\right]f_\vartheta^\mathcal{O}(\theta_3,\dots,\theta_n).
  \end{align}
\end{itemize}
These three axioms give information about how the form factors behave under the exchange of two particles in the state, under the application of crossing symmetry, and under the annihilation of particle-hole excitation, respectively. The axioms depend on two functions: the  scattering matrix $S(\theta_1, \theta_2)$ and $R_{\vartheta}(\theta|\theta_1, \dots, \theta_n)$ which describes the collective effect of the background particles on the scattering of the particle with rapidity $\theta$ due to the presence of the additional particles $\theta_1, \dots, \theta_n$. This function describes the back-reaction of the background upon inserting extra particles. In the low density limit the $R_{\vartheta}$ function tends to $1$ and the axioms turn into the standard vacuum form-factors axioms.

We finish the introduction with a few important remarks.

First, we point out that the expansion (\ref{lehmann}) in terms of form factors, is not a low-temperature (or low-energy-density) expansion. The expansion is centered around the representative state, $\vert \vartheta\rangle$, and the leading contributions to (\ref{lehmann}) will come from the cases where  the states $\vert \vartheta^\prime\rangle$ are small deviations around $\vert \vartheta\rangle$. The state $\vert \vartheta\rangle$ is far from the vacuum, and there is no restriction for the energy density to be low.

Second, an approach for the computation of  correlation functions similar to the one we are describing here has already proven to be successful in the study of correlation functions of the non-relativistic Lieb Liniger model \cite{1742-5468-2015-2-P02019,SciPostPhys.1.2.015,2018JSMTE..03.3102D}. In this model, form factors with an arbitrary number of particle excitations are known from using algebraic Bethe ansatz techniques \cite{korepin_bogoliubov_izergin_1993}. The form factors of the form $\langle \vartheta^\prime\vert\mathcal{O}(0)\vert\vartheta\rangle$ can then be computed by taking the thermodynamic limit of the many-particle form factors. Considering QFT's instead of the Lieb Liniger model will give us a  set of additional analytic tools, since relativistic invariance places very strong restrictions on  such form factors, and the form factors will be computed by finding functions that satisfy all these restrictions. Our approach will then be very different than that used for the Lieb Liniger model in \cite{1742-5468-2015-2-P02019}. Our approach {\it will not} be to take the thermodynamic limit of a previously known many-particle form factor (which are extremely difficult to calculate exactly in QFT). Instead, we will derive the finite-energy-density form factors by finding functions which solve the large number of constraints which arise from integrability and relativistic invariance in the presence of the finite-energy-density background, without ever having to compute complicated many-particle form factors.

Third, we point out that the idea of computing form factors in the presence of a finite-energy density background in general is not new. In particular in~\cite{Doyon:2005jf}, an approach was developed for Ising field theory for computing form factors with excitations on top of the thermal Gibbs ensemble. The authors dervied a set of axioms for local operators (for which the new axioms are simple in this non-interacting theory), as well as non-trivial semilocal operators (which we will not discuss in this paper). It was however, not easy at all to generalize this approach to  interacting field theories. The reason this approach is so much more difficult than ours, is that in \cite{Doyon:2005jf}, it was attempted to compute form factors including excitations on top of the genuine Gibbs ensemble, which involves a sum over an infinite number of states. Our approach is made much simpler by using the concept of the representative state (\ref{representativestate}). It is conceptually much simpler to understand what it means to have an excitation over the representative state, while adding an excitation on top of an averaging ensemble is a much more elusive concept, making it much more difficult to handle the form factors and derive axioms for them.

The organization  of the manuscript is the following. In Sections~\ref{sec:bootstrap_vacuum} and~\ref{sec:QFT_finite_volume} we introduce the neccessary background on the vacuum form factor bootrap program and on  integrable QFT's at finite volume. Some standard computations of the thermodynamic limit of the finite volume QFT are recalled in Appendix~\ref{app:finite_volume}.  Those two sections (and Appendix) serve as a basis for Section~\ref{sec:bootstrap_background}, where we formulate the axioms satisfied by the finite-background form factors. The logic behind the derivation of these axioms  is the following. We start with knowledge of the vacuum form-factor axioms in the infinite volume. We then place the system in a finite volume, and use previously known relations between the finite-and infinite-volume form factors \cite{Pozsgay:2007kn}. We use these relations to derive constraints satisfied by finite-volume form factors, with a large, but finite number of particles. Then we go back to the infinite volume limit but this time we keep the density of particles fixed. In taking this thermodynamic limit, the relations we found for the finite-volume form factors yield the new axioms for the finite-background form factors, now at infinite volume.

 After deriving the axioms, in Section~\ref{sec:consequences} we analyze the consequences they impose on the form factors. Specifically we introduce a notion, familiar from the vacuum form factors axioms, of minimal form-factors. In Section~\ref{sec:corr_func} we conjecture a formula for the correlation function which resembles the well-known LeClair-Mussardo formula. The main difference lies in the form factors used, which in our case are now dressed by the finite background. Our two-point function is also already regularized, since the problems with kinematical poles are already absorbed into the definition of the finite-background form factors.  In the Section~\ref{sec:infrared} we show that there are generally two distinct infrared limits of the correlation functions, which show either exponential or polynomial decay at long distances. The closing section~\ref{sec:SinhGordon} is devoted to Sinh-Gordon theory, a simple example of an integrable field theory. We show there, in a concrete example, how the form factor axioms can be used to compute the one-and two-particle form factors of vertex operators in the presence of a thermal background.

\section{The form factor bootstrap program}\label{sec:bootstrap_vacuum}

We first review the standard bootstrap approach to form factors in integrable QFT. We assume diagonal scattering for simplicity, and consider theories with one species of particle. We parametrize the energy and momentum of a particle of mass $m$, as $E=m\cosh\theta,\,p=m\sinh\theta$, respectively, where $\theta$ is the particle's rapidity.

We define the $n$-particle form factor of some local operator, $\mathcal{O}(x)$, as
\beq
f^{\mathcal{O}}(\theta_1,\dots,\theta_n)\equiv\langle 0\vert \mathcal{O}(0)\vert \theta_1,\dots,\theta_n\rangle.\nonumber
\eeq
ordered as $\theta_1>\theta_2>\dots>\theta_n$. We can also consider matrix elements with particles in the outgoing state, using the crossing relation~\cite{doi:10.1142/1115}
\beq
f^{\mathcal{O}}(\theta^\prime\vert \theta_1,\dots,\theta_n)&\equiv&\langle\theta^\prime\vert \mathcal{O}(0)\vert\theta_1,\dots,\theta_n\rangle\nonumber\\
&=& \left[\sum_{k=1}^n2\pi\delta(\theta^\prime-\theta_k)\prod_{l=1}^{k-1} S(\theta_l-\theta_k)f^{\mathcal{O}}(\theta_1,\dots,\theta_{k-1},\theta_{k+1},\dots,\theta_n)\right]\nonumber\\
&&+f^{\mathcal{O}}(\theta^\prime+\pi{\rm i},\theta_1,\dots,\theta_n),\label{generalizedcrossing}
\eeq
where $S(\theta_i-\theta_j)$ is the two-particle S-matrix, which depends only on the difference of the two rapidities, as a consequence of Lorentz invariance. The last term in the right hand side of (\ref{generalizedcrossing}) is the connected piece of the form factor.

The form factors satisfy a set of very restricting axioms, which are a consequence of elasticity and factorization in the integrable theory, as well as other physical considerations, such as Lorentz invariance, and unitarity. An exact expression for the form factors can often be computed by finding a solution to the set of axioms. We now list all these restrictions.

\begin{enumerate}

\item {\it Lorentz symmetry}

If $\mathcal{O}(x)$ is an operator with Lorentz spin, $s$, then under a  boost, where all the rapidities are shifted as $\theta_i\to\theta_i+\alpha$, the form factor transforms as 
\beq
f^{\mathcal{O}}(\theta_1+\alpha,\dots,\theta_n+\alpha)=e^{s\alpha}f^{\mathcal{O}}(\theta_1,\dots,\theta_n).\nonumber
\eeq
In particular, if $\mathcal{O}(x)$ is a scalar operator, with $s=0$, then it is invariant under a Lorentz boost. The Lorentz symmetry implies, that besides the $s$-dependent pre-factor, the form factors depend only on the differences of rapidities $\theta_{ij}=\theta_i-\theta_j$. Therefore the form factor depends independently only on $n-1$ of the $n$ rapidities.

\item{\it Scattering axiom}

The order of two particles in the incoming state may be exchanged by scattering the two particles, as
\beq
f^{\mathcal{O}}(\theta_1,\dots,\theta_i,\theta_{i+1},\dots,\theta_n)=S(\theta_{i}-\theta_{i+1})f^{\mathcal{O}}(\theta_1,\dots,\theta_{i+1},\theta_i,\dots,\theta_n),\label{scatteringaxiom}
\eeq
where $S(\theta_i-\theta_{j})$ is the two-particle S-matrix. The S-matrix can typically be derived from a set of axioms which follow from integrability, we refer the reader to \cite{ZAMOLODCHIKOV1979253,Dorey:1996gd} for more information. Here, we assume the S-matrix of the IQFT is known, and use it as an input in the form factor bootstrap program. 

\begin{figure}
  \center
  \includegraphics[scale=0.5]{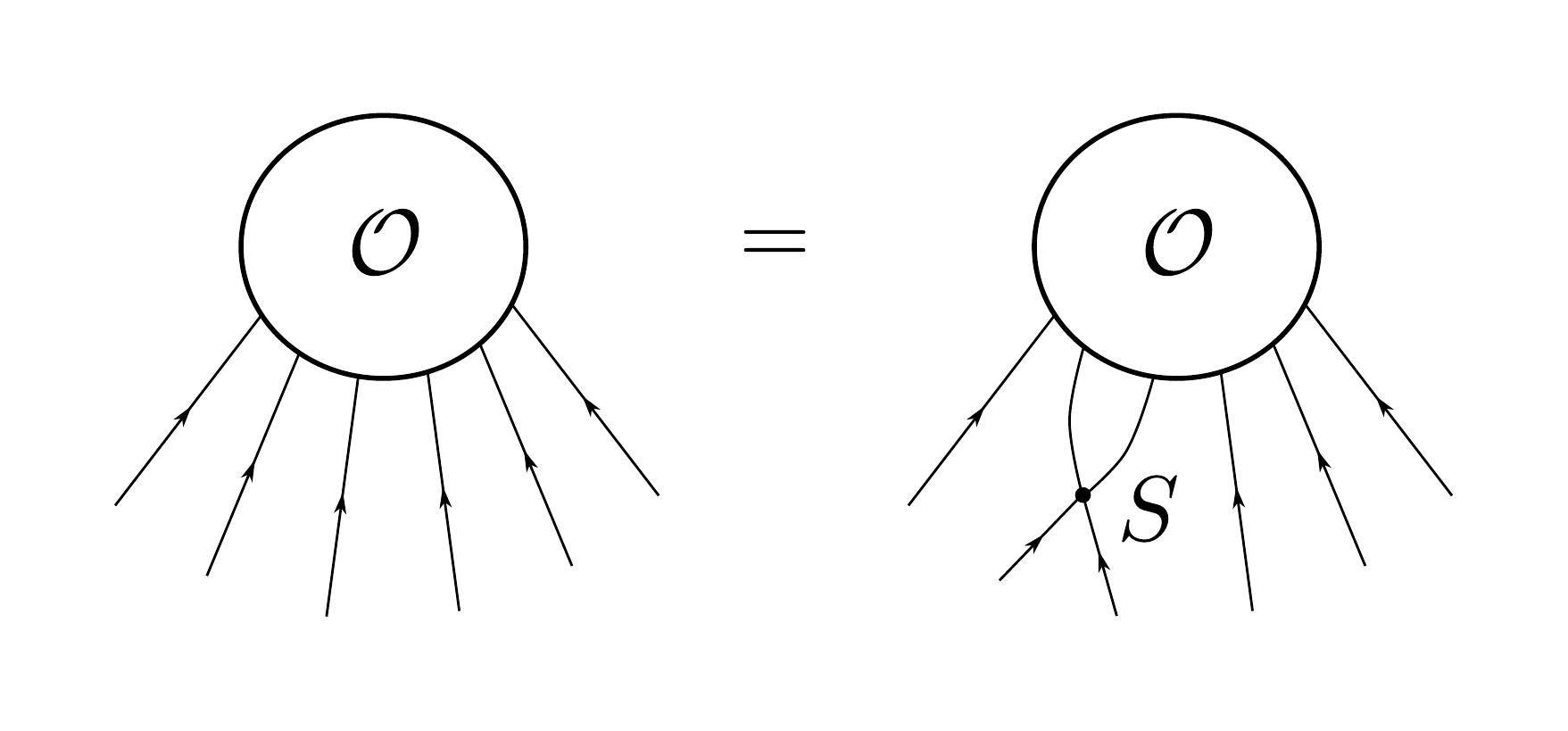}
  \caption{Graphical representation of the scattering axiom}\label{fig:scattering_vacuum}
\end{figure}

\item{\it Periodicity axiom}

Following the generalized crossing relation, (\ref{generalizedcrossing}), a particle in the incoming state may be turned into an outgoing particle by shifting its rapidity by $\pi {\rm i}$. By applying crossing symmetry twice, we find that we can turn the last particle in the incoming state, into the first one,  as
\beq
f^{\mathcal{O}}(\theta_n+2\pi,\theta_1,\dots,\theta_{n-1})=f^{\mathcal{O}}(\theta_1,\dots,\theta_n).\label{periodicityaxiom}
\eeq
This axiom may be modified when the particle $\theta_n$ and the operator $\mathcal{O}(x)$ are not local with respect to each other (for example when the particle is a topological soliton), but for simplicity, we will not consider such cases for now.

\begin{figure}
  \center
  \includegraphics[scale=0.5]{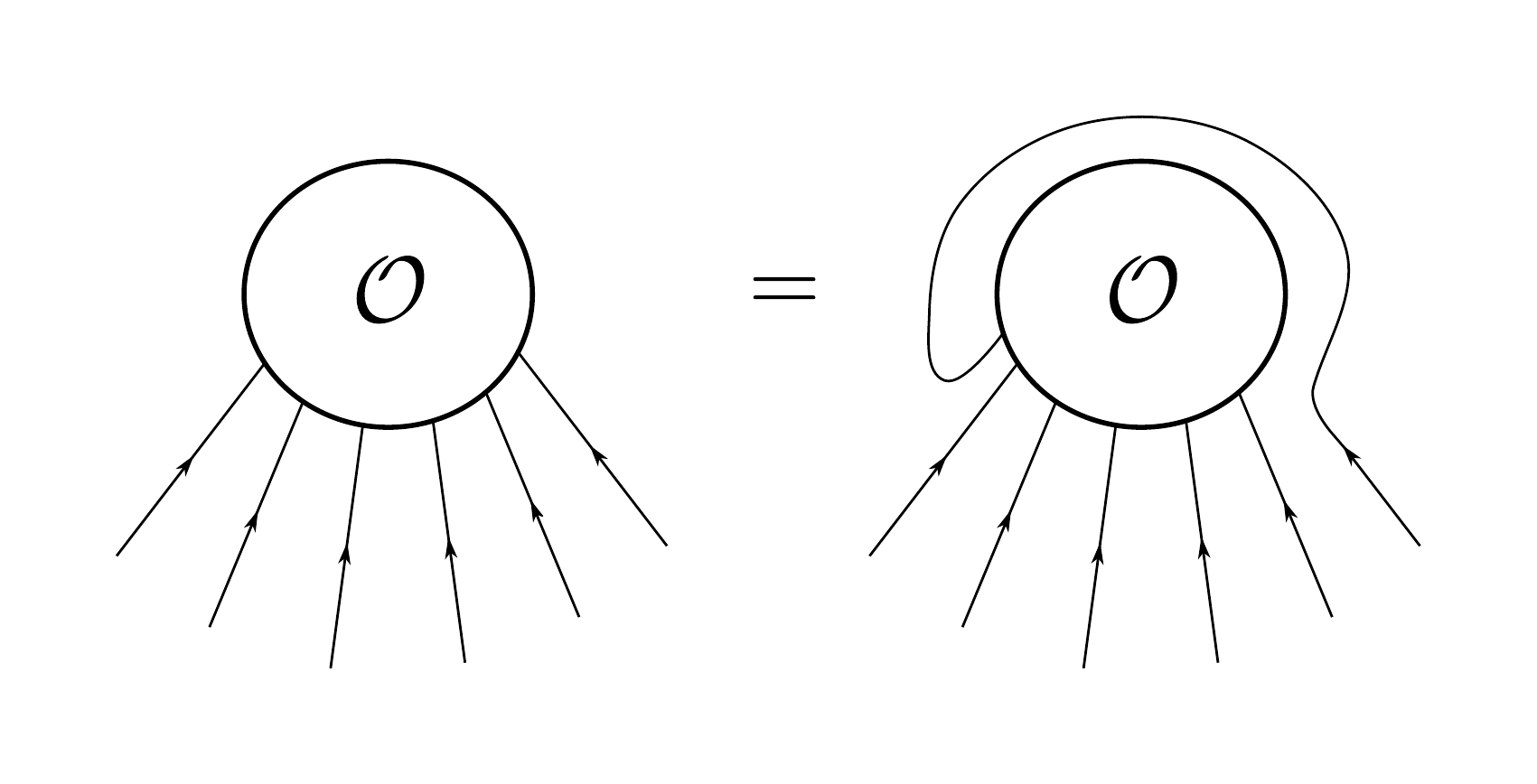}
  \caption{Graphical representation of the periodicity axiom}\label{fig:periodicity_vacuum}
\end{figure}

The three axioms presented so far are enough to compute the so-called ``minimal form factors", 
\beq
f^{\rm min}(\theta_1,\dots,\theta_n)=\prod_{i<j}f^{\rm min}(\theta_{ij}),\nonumber
\eeq
which are independent of the particular operator, $\mathcal{O}$. These minimal form factors are defined as the maximally analytic solutions of the scattering and periodicity axioms, {\it i.e.} satisfying the so-called ``monodromy properties",
\beq
S(\theta_{ij})f^{\rm min}(-\theta_{ij})=f^{\rm min}(2\pi{\rm i}-\theta_{ij}),\label{monodromy}
\eeq
with no poles or zeroes in the ``physical region", $\theta_{ij}\in [0,2\pi {\rm i}]$.

\begin{figure}
  \center
  \includegraphics[scale=0.5]{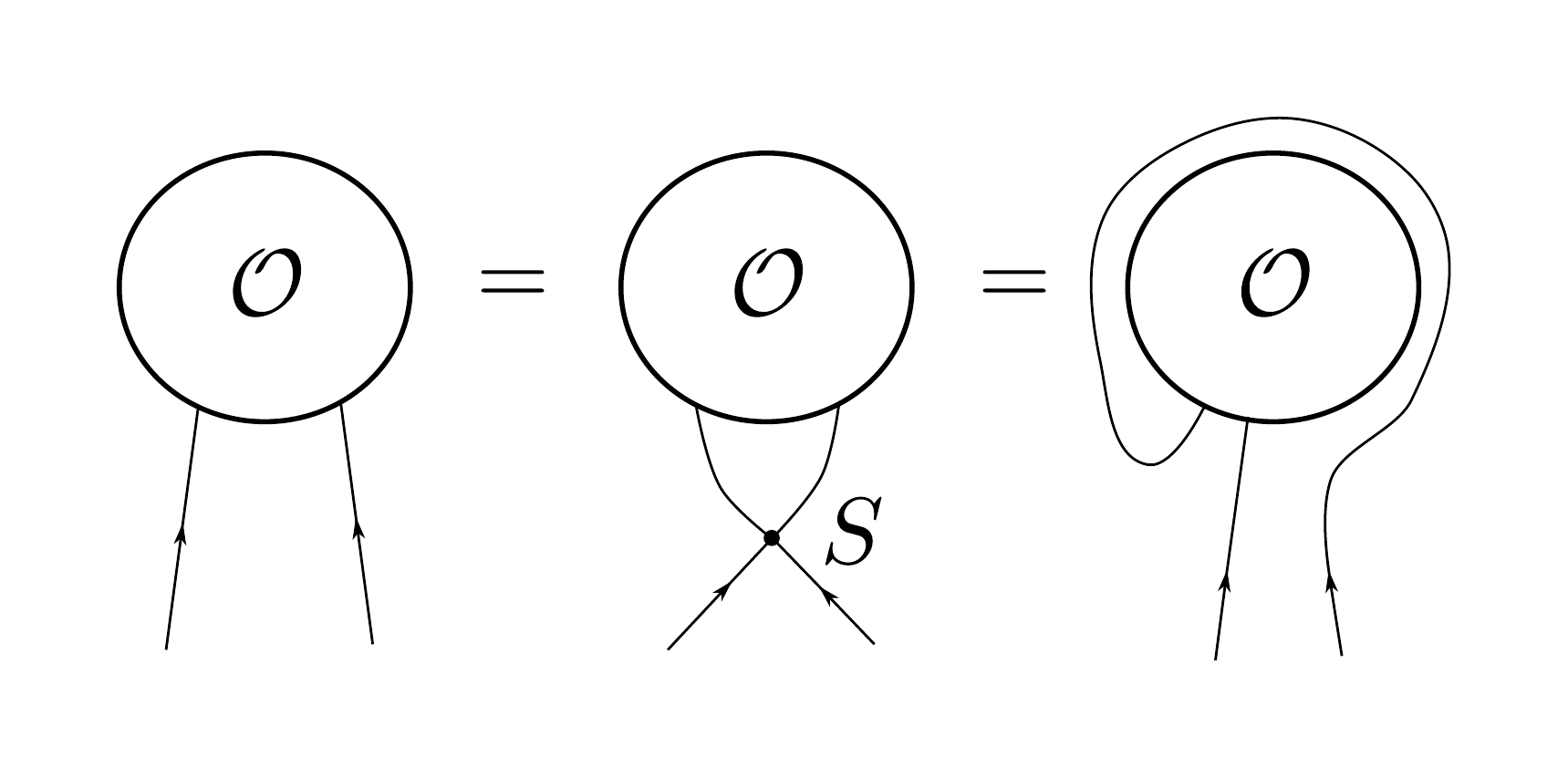}
  \caption{Graphical representation of the monodromy property}\label{fig:minimal_vacuum}
\end{figure}

The minimal solution of (\ref{monodromy}) can be found~\cite{KAROWSKI1978455} by expressing the S-matrix as
\beq
S(\theta)=\exp\left[\frac{1}{2}\int_{-\infty}^\infty\frac{dt}{t}B(t) \exp\frac{t\theta}{\rm i\pi}\right].\nonumber
\eeq
Then the minimal form factor is
\beq
f^{\rm min}(\theta)=\exp\left[-\frac{1}{4}\int_{-\infty}^\infty\frac{dt}{t}\frac{B(t)}{\sinh t}\exp\left(\frac{t(\theta - i\pi)}{i\pi}\right)\right].\label{fmin}
\eeq

The general form factors can be expessed as 
\beq
f^{\mathcal{O}}(\theta_1,\dots,\theta_n)=K^{\mathcal{O}}(\theta_1,\dots,\theta_n)f^{\rm min}(\theta_1,\dots,\theta_n),\nonumber
\eeq
where $K^{\mathcal{O}}$ is necessarily a periodic function in all the rapidities,
\beq
K^{\mathcal{O}}(\theta_1,\dots,\theta_i+2\pi {\rm i},\dots,\theta_n)=K^{\mathcal{O}}(\theta_1,\dots,\theta_i,\dots,\theta_n),\nonumber
\eeq
for any $i=1,\dots,n$. These functions $K^\mathcal{O}$ must contain all necessary poles in the physical region, and any dependence on the particular operator, $\mathcal{O}(x)$.  These function can be expressed generally as
\beq
K^{\mathcal{O}}(\theta_1,\dots,\theta_n)=\frac{Q^{\mathcal{O}}(\theta_1,\dots,\theta_n)}{D(\theta_1,\dots,\theta_n)},\nonumber
\eeq
where $Q^{\mathcal{O}}(\theta_1,\dots,\theta_n)$ and $D(\theta_1,\dots,\theta_n)$ can be chosen to be  polynomials involving powers of $\cosh(\theta_{ij})$. The denominator, $D(\theta_1,\dots,\theta_n)$ is chosen as the minimal function that includes all the necessary poles in the physical region, and does not depend on the operator, $\mathcal{O}(x)$. All the operator dependence is then incorporated in the numerator function, $Q^{\mathcal{O}}(\theta_1,\dots,\theta_n)$.

\item {\it Annihilation-pole axiom}

In a relativistic QFT, it is possible for a particle in the incoming state to annihilate with its antiparticle on the outgoing state (in the simplest cases we consider, the particle is its own antiparticle). As a consequence, the form factor has a kinematical annihilation pole, for example for the particles with rapidities $\theta_1$ and $\theta_2$, at the point $\theta_1=\theta_2+\pi {\rm i}$. The annihilation-pole axiom relates the residue of the $n$-particle form factor at the annihilation pole, to the $n-2$ particle form factor, where the two particles have annihilated, explicitly,
\beq
-{\rm i}{\rm Res}_{\theta_1=\theta_2}f^\mathcal{O}(\theta_1+\pi {\rm i},\theta_2,\theta_3,\dots,\theta_n)=\left(1-\prod_{k=3}^n S(\theta_{2} -\theta_k)\right)f^\mathcal{O}(\theta_3,\dots,\theta_n).\label{annihilationpoleaxiom}
\eeq
This axiom then provides a recursive relation, linking the $n$ and $n-2$ particle form factors. This recursion can often be used to normalize higher-particle form factors, based on the normalization of lower form factors.

\item{\it Cluster properties}

If one boosts the rapidities of only a subset of the particles in the incoming state, the form factors are expected to satisfy the so-called ``cluster properties"~\cite{1996PhLB..387..327D},
\beq
\lim_{\alpha\to\infty} f^{\mathcal{O}_1}(\theta_1,\dots,\theta_i,\theta_{i+1}-\alpha,\dots,\theta_n-\alpha)=\mathcal{N}_{{\rm cluster}}^{\mathcal{O}_1}f^{\mathcal{O}_2}(\theta_1,\dots,\theta_i)f^{\mathcal{O}_3}(\theta_{i+1},\dots,\theta_n),\nonumber
\eeq
where $\mathcal{N}_{\rm cluster}^{\mathcal{O}_1}$ is some normalization constant, and $\mathcal{O}_{2,3}$ are some other local operators.  In particular, this relation is expected to simplify in simple theories with no additional internal symmetries \cite{Acerbi:1996zj}, where some of the form factors should ``self cluster", such that $\mathcal{O}_1=\mathcal{O}_2=\mathcal{O}_3$, and $\mathcal{N}_{\rm cluster}^{\mathcal{O}_1}=1/\langle 0\vert \mathcal{O}_1\vert 0\rangle$.

\item {\it Bounds on growth}

One can further restrict the function $Q^\mathcal{O}(\theta_1,\dots,\theta_n)$ by placing bounds on how fast this function can grow as we take $\vert \theta_i\vert\to\infty$. An upper bound on the growth can be derived (for unitary theories) by considering the two-point function of the operator $\mathcal{O}$, which at very short distances can be described by an underlying CFT, using an operator product expansion. The derivation can be found in \cite{1995NuPhB.455..724D,mussardobook}, and here we only cite the result,
\beq
\lim_{\vert\theta_i\vert\to\infty}f^\mathcal{O}(\theta_1,\dots,\theta_n)\sim e^{y_{\mathcal{O}}\vert\theta_i\vert},\label{growthconstraint}
\eeq
with $y_\mathcal{O}\leq \Delta_\mathcal{O}$, where $\Delta_{\mathcal{O}}$ is the conformal weight of the operator.  This constraint (\ref{growthconstraint}) places an upper bound on the degree of the polynomial, $Q^\mathcal{O}(\theta_1,\dots,\theta_n)$, and this bound is specific to the particular operator.

\item{\it Translation}

Once the form factor of the operator at the origin, $x=0$, is known, it is easy to obtain the form factor for the operator at any space-time point, $x$. Using Lorentz symmetry, one can write
\beq
f^{\mathcal{O}}(x,t)(\theta_1,\dots,\theta_n)=e^{{\rm i}xP_n- {\rm i}tE_n}f^\mathcal{O}(\theta_1,\dots,\theta_n),\nonumber
\eeq
where $P_n=\sum_{i=1}^n m\sinh\theta$, and $E_{n}=\sum_{i=1}^n m\cosh\theta$ are the total momentum and energy, respectively, of the $n$-particle state.

\end{enumerate}

In addition to these axioms, for theories whose spectrum includes particle bound states, there exist additional axioms, concerning the additional bound-state pole structure. If the the interactions between particles are such that they may form bound states, the form factors will have additional poles in the physical region, corresponding to the fusion angle between the two elementary particles. It is also possible for the form factor to have higher-order poles corresponding to processess with bound states as intermediate states \cite{doi:10.1142/1115}. For simplicity we will not consider theories with bound states here. For simplicity, we consider only QFT's with no bound states, and only one particle species, so we will not list any bound-state axioms, or their modification at finite background.

We note also that we are only considering operators, $\mathcal{O}(x)$, which are {\it local}, relative the particle excitations. This means we are excluding the computation of semi-local operators, which can also be studied in the bootstrap approach by modifying some of the form factors with some semi-locality factor. Semi-local operators commonly arise, for example, in cases where the particle excitations are topological kinks (such as in the Ising field theory, or the sine-Gordon model), and are therefore not purely local excitations of the fundamental field. The derivation of finite-background form factor axioms for semi-local operators is a much harder problem, since there are fundamental problems with the definition of semi-local operators at finite volume (which we need to formulate as an intermediate state before taking the thermodynamic limit), which induce non-trivial effects, such as operator mixing. These difficulties were encountered and discussed for the non-interacting Ising field theory in \cite{Doyon:2005jf}, but they are beyond the scope of this paper.

\section{Integrable QFT in a finite volume}\label{sec:QFT_finite_volume}

Now we want to consider form factors of local operators with incoming and outgoing states with a finite  energy density background, and finite excitations on top of this background. This background can be defined by first placing the system at finite volume, $L$, and then taking the thermodynamic limit, $L, n\to\infty$, with the particle density, $n/L$, fixed. We then first need to understand how to work with form factors at finite volume. Most of the background knowledge introduced in this section can be found in more detail in the original references \cite{ZAMOLODCHIKOV1990695} (the first use of the thermodynamic Bethe ansatz for QFT), and \cite{Pozsgay:2007kn} (derivation of expressions for finite-volume form factors).

We place system at finite volume and assume periodic (anti-periodic) boundary conditions for bosons (fermions). This introduces a quantization of allowed particle rapidities. We shall be mainly concerned with the integrable theories of the \emph{fermionic type}\footnote{It is known that all {\it interacting} integrable QFT's with diagonal scattering are of the fermionic type. The only known bosonic type theory is the free boson, for which the form factor program is trivial, so for our purposes considering only fermionic-type theories is sufficient.} in which
\beq
S(0) = -1.
\eeq
 The quantization conditions are given by the Bethe equations, 
\beq
e^{i m L \sinh\tilde{\theta}_k} \prod_{l \neq k} S(\tilde{\theta}_k - \tilde{\theta}_j) = 1,\qquad  k=1,\dots, n.
\eeq
It is convenient to consider the logarithm of this expression, to find the allowed rapidities for an $n$-particle state satisfy
\beq
Q_k(\tilde{\theta}_1,\dots,\tilde{\theta}_n) \equiv mL\sinh\tilde{\theta}_k+\sum_{l\neq k}\delta_0(\tilde{\theta}_k-\tilde{\theta}_l)=2\pi I_k,
\label{bethequantization}\eeq
where $\delta_0(\theta)\equiv -{\rm i}\log (-S(\theta))$, $I_k$ is a quantum number, and there is one such equation for each $k=1,\dots,n$.
For fermionic (antiperiodic) boundary conditions, the quantum numbers are integers, while for bosonic (periodic) boundary conditions they are integers when $n$ is odd and half-odd integers for $n$ even.
The tildes on the rapidity variables denote that these are finite-volume solutions of the quantization equations (\ref{bethequantization}). A state is then fixed by choosing a set of integers, $\{I_k\}$, which are an input for finding the solutions $\{\tilde{\theta}_k\}$ of (\ref{bethequantization}). The total energy of the $n$-particle state is given by 
\beq
E=\sum_{i=1}^n m\cosh\tilde{\theta}_i.\nonumber
\eeq
At finite volume, it is then convenient to switch to an ``integer" basis of states, such that
\beq
\vert I_1,\dots,I_n\rangle=\frac{1}{\sqrt{\rho(\tilde{\theta}_1,\dots,\tilde{\theta}_n)}}\vert \tilde{\theta}_1,\dots,\tilde{\theta}_n\rangle,\nonumber
\eeq
where $\rho(\tilde{\theta}_1,\dots,\tilde{\theta}_n)$ is the determinant of the Jacobian of the transformation between the two bases,
\beq
J_{kl}(\tilde{\theta}_1,\dots,\tilde{\theta}_n)&=&\frac{\partial}{\partial\tilde{\theta}_l}Q_k(\tilde{\theta}_1,\dots,\tilde{\theta}_n),\nonumber\\
\rho(\tilde{\theta}_1,\dots,\tilde{\theta}_n)&=& |{\rm Det} \,J(\tilde{\theta}_1,\dots,\tilde{\theta}_n)|.\nonumber
\eeq
We point out that these determinants, $\rho(\{\tilde{\theta}\})$ are periodic under the transformations $\tilde{\theta}_i\to\tilde{\theta}_i+2\pi{\rm i}$. They are also symmetric functions of $\tilde{\theta}_j$. 

The general form factors with particles in the incoming and outgoing states, without any disconnected pieces is given by~\cite{Pozsgay:2007kn}
\beq
\langle I^\prime_1,\dots,I_m^\prime\vert \mathcal{O}(0)\vert I_1,\dots,I_n\rangle=\frac{f^\mathcal{O}(\tilde{\theta}_1^\prime+\pi{\rm i},\dots,\tilde{\theta}_m^\prime+\pi{\rm i},\tilde{\theta}_1,\dots,\tilde{\theta}_n)}{\sqrt{\rho(\tilde{\theta}_1^\prime,\dots,\tilde{\theta}_m^\prime)\rho(\tilde{\theta}_1,\dots,\tilde{\theta}_n)}}+O\left(e^{-\mu L}\right).\label{finitevolumeff}
\eeq
The expression (\ref{finitevolumeff}) is exact at finite volume to all orders of powers of $1/L$ corrections, and the only corrections are exponentially suppressed. These corrections, however interesting for the structure of the finite volume IQFT~\cite{2018JHEP...07..174B}, do not play any role in the thermodynamic limit. It is important to note that the only way the form factor (\ref{finitevolumeff}) can have any disconnected contributions, is if the two sets of rapidities in the incoming and outgoing states are exactly equal to each other, or $\{I^\prime\}=\{I\}$, this is because changing one of the integers, will produce an $O(1/L)$ shift in all of the other rapidities, $\tilde{\theta}_i$,  which means the rapidities in the incoming and outgoing states will no longer match.

The thermodynamic limit is taken by increasing $L$ while leaving the energy density constant. The total energy of the finite-density state in the thermodynamic limit can be written as \cite{ZAMOLODCHIKOV1990695}
\beq
E=L\int d\theta\rho_p(\theta) \,m\cosh\theta,\nonumber
\eeq
where $\rho_p(\theta)$ is a distribution describing the density of particles. We can also define the density of available states, $\rho_s(\theta)$, with which we define the particle occupation number $\vartheta(\theta)=\rho_p(\theta)/\rho_s(\theta)$. The quantization conditions in the thermodynamic limit become an integral equation for the densites (c.f. Appendix~\ref{app:finite_volume})
\beq
2\pi \rho_s(\theta)=m\cosh\theta+\int d\theta^\prime\varphi(\theta-\theta^\prime)\rho_p(\theta^\prime),\label{tba_rho}
\eeq
so that only one of the distributions $\rho_p(\theta),\,\rho_s(\theta),\,\vartheta(\theta)$ is independent and contains all macroscopic information. We choose $\vartheta(\theta)$ to play this role. In the formula above we have defined $\varphi(\theta)=\partial \delta(\theta)/\partial\theta$. We also note that function $\rho_s(\theta)$ is $2\pi i$ periodic. It will later be convenient to also define a density of holes, $\rho_h(\theta)$, or density of available states which are not occupied by particles, such that $\rho_s(\theta)=\rho_p(\theta)+\rho_h(\theta)$.

Acquired now with the tools to describe the thermodynamic limit we consider this limit for a number of quantities. 
We define, in the thermodynamic limit, a state corresponding to some macroscopic distribution, $\vartheta(\theta)$, as
\beq
\vert \vartheta\rangle= \exp\left(-\frac{1}{2}S[\vartheta]\right)\lim_{\rm th} \sum_{\{\tilde{\theta}\}}\vert \{\tilde{\theta}\}\rangle,\label{state_TL}
\eeq
where one sums over all the microscopic states with particle content $\{\tilde{\theta}\}$, leading to the same macroscopic distribution, $\vartheta$. $S[\vartheta]$ is the Yang-Yang entropy associated with this state,
\beq
S[\vartheta] = L\int_{-\infty}^{\infty} d\theta s[\vartheta,\theta], \qquad s[\vartheta,\theta] = \rho_s(\theta)\log \rho_s(\theta) - \rho_p(\theta)\log \rho_p(\theta) - \rho_h(\theta)\log \rho_h(\theta).\label{defs}
\eeq
The norm of the state $\vert \vartheta\rangle$ is 
\beq
 \langle \vartheta \vert \vartheta\rangle = \exp\left(-S[\vartheta]\right)\lim_{\rm th} \sum_{\{\tilde{\theta}\}}\rho(\{\tilde{\theta}\}),
\eeq
given that the finite-volume states are normalized as $\langle \{I\}\vert\{I\}\rangle=1$. The thermodynamic limit on the right hand side is not well-defined because functions $\rho(\{\tilde{\theta}\})$ contain some irregular terms~\cite{2011_Shashi_PRB_84}. If we consider a normalized form factor, the irregular terms cancel with corresponding terms in the form factor. Therefore, only the normalized form factor has a well-defined thermodynamic limit. For this reason the right hand side of this formula should be understood as an intermediate expression. We will formulate the axioms for normalized form factors and therefore we shall not be concerned with giving precise description of this thermodynamic limit.

Introducing the additional particles induces a change in the rapidities of the background. This back-reaction of the background particles 
\beq
\tilde{\theta}\to \tilde{\theta} - \frac{F(\tilde{\theta}\vert\{\theta_i\})}{L \rho_s(\theta) },\nonumber
\eeq
is captured by the back-flow function $F(\tilde{\theta}\vert\{\theta_i\})$. We assume that the back-flow function is separable. The full back-flow can be written as a sum of separate contributions coming from each particle $\{\theta_i\}$,
\beq
F(\tilde{\theta}\vert \{\theta_i\})=\sum_i  F(\tilde{\theta}\vert \theta_i),\nonumber
\eeq
The shift in the background rapidities has two origins. First, it is the effect of interactions between the particles. Second, in bosonic theory with periodic boundary conditions, there is an additional shift. This additional shift occurs to keep the boundary conditions periodic. When adding or removing a particle, the quantum numbers must be shifted by $1/2$. This results in an extra shift of rapidities. We shall call the back-flow function which takes into account only the shift due to the interactions, the fermionic back-flow function and denote it $F_F(\tilde{\theta}\vert\{\theta_i\})$. The bosonic back-flow function $F_B(\tilde{\theta}\vert\{\theta_i\})$ takes into account both effects.

The single particle fermionic back-flow is given by (c.f. Appendix~\ref{app:finite_volume})
\beq
 2\pi F_F(\tilde{\theta}|\theta_1) = \delta_F(\tilde{\theta} - \theta_1) + \int d\theta' \vartheta(\theta') \varphi(\tilde{\theta} - \theta')F_F(\theta'| \theta_1), \qquad \delta_F(\theta) = \delta_0(\theta).
\eeq 
We find the hole back-flow to be minus back-flow of the particle,
\beq
F_F(\tilde{\theta}|\theta_1 + i\pi) = - F_F(\tilde{\theta}|\theta_1). 
\eeq
We will introduce the bosonic back-flow function later. Let us now turn to the problem of defining a finite-volume representation of a state with excitations over the background.

We denote the state specified by a particle filling function $\vartheta(\theta)$ and a set of additional excitation $\{\theta_i\}$ by $\vert\vartheta, \theta_1, \dots, \theta_n\rangle$. This state is understood as thermodynamic limit of the finite-volume state
\beq
  \vert\vartheta, \theta_1, \dots, \theta_n\rangle &=& \exp\left(-\frac{1}{2}S[\vartheta; \theta_1, \dots, \theta_n]\right) \lim_{\rm th}   \sum_{\{\tilde{\theta}^\prime\}}\vert \{\tilde{\theta}^\prime\}, \tilde{\theta}_1,\dots,\tilde{\theta}_n\rangle,\nonumber\\
  &=& \exp\left(-\frac{1}{2}S[\vartheta;\theta_1,\dots,\theta_n]\right)\lim_{\rm th}\sum_{\{I\}} \sqrt{\rho(\{\tilde{\theta}^\prime\},\tilde{\theta}_1,\dots,\tilde{\theta}_n)}\,\,\vert \{I\},I_1,\dots,I_n\rangle\label{state_TL2}
\eeq
where again one sums over all the microscopic states leading to the same macroscopic state. The prime in $\{\tilde{\theta}^\prime\}$ indicates that the rapidities in the background state are modified by the back-flow function. At this stage it is convenient to split the quantum numbers in two sets: $\{I\}$ is a set of integers which leads to the distribution $\vartheta(\theta)$ in the thermodynamic limit, and $I_1,\dots,I_n$ are the integers corresponding to the additional excitations. The choice of $\{I\}$ is not unique, indeed there are many choices of a quantum number that in the thermodynamic limit yield the same excitation $\theta$, which is why we average by summing over all possible such sets $\{I\}$.

We are interested in form factors of the form $\langle \vartheta|\mathcal{O}|\vartheta, \theta_1, \dots, \theta_n\rangle$. We relate them to the thermodynamic limit of the finite system form factors in the following way. From the definition~\eqref{state_TL} and~\eqref{state_TL2} it follows that
\beq
  \langle \vartheta|\mathcal{O}|\vartheta, \theta_1, \dots, \theta_n\rangle &=& \exp\left(-\frac{1}{2}\left(S[\vartheta] + S[\vartheta, \theta_1, \dots, \theta_n] \right) \right) \lim_{\rm th}  \sum_{\{\tilde{\theta}\}, \{\tilde{\theta}'\}} \langle\{\tilde{\theta}\}\vert\mathcal{O}\vert\{\tilde{\theta}'\}, \tilde{\theta}_1,\dots,\tilde{\theta}_n\rangle. \nonumber\\
  &=&\exp\left(-S[\vartheta] - \frac{1}{2}\delta S[\vartheta, \theta_1, \dots, \theta_n] \right)  \lim_{\rm th}  \sum_{\{\tilde{\theta}\}, \{\tilde{\theta}'\}} \langle\{\tilde{\theta}\}\vert\mathcal{O}\vert\{\tilde{\theta}'\}, \tilde{\theta}_1,\dots,\tilde{\theta}_n\rangle. \label{derivation_ff_tl}
\eeq
where function $\delta S[\vartheta,\{\theta_i\}]$ is the differential entropy~\cite{1990_Korepin_NPB_340}
\beq
\delta S[\vartheta, \theta_1] = \int_{-\infty}^{\infty} d\theta' s[\vartheta, \theta'] \frac{\partial}{\partial \theta'} \left( \frac{F(\theta', \theta_1)}{\rho_s(\theta')}\right),
\eeq
with $s[\vartheta, \theta']$ defined in~\eqref{defs}. The differential entropy is antiperiodic in $\theta_1$,
\beq
\delta S[\vartheta, \theta_1 + i\pi] = - \delta S[\vartheta, \theta_1].
\eeq

To evaluate the sum in~\eqref{derivation_ff_tl} we follow the prescription introduced in~\cite{QA_Caux}. The main point is that the form factors $\langle\{\tilde{\theta}\}\vert\mathcal{O}\vert\{\tilde{\theta}'\}, \tilde{\theta}_1,\dots,\tilde{\theta}_n\rangle$ in the thermodynamic limit are almost diagonal in sets $\{\tilde{\theta}\}$ and $\{\tilde{\theta}'\}$. Therefore, choice of $\{\tilde{\theta}\}$ fixes $\{\tilde{\theta}'\}$ up to small (of order $1/L$) displacements. These displacements correspond to tiny non-dispersive excitations, referred to as soft modes. The summation over the soft modes leads to a renormalized, dressed, form factor
  \begin{equation}
    \langle\{\tilde{\theta}\}\vert\mathcal{O}\vert\{\tilde{\theta^\prime}\}, \tilde{\theta}_1,\dots,\tilde{\theta}_n\rangle_{\rm dressed} = \sum_{{\rm soft\, modes}} \langle\{\tilde{\theta}\}\vert\mathcal{O}\vert\{\tilde{\theta^\prime} + {\rm soft\, modes}\}, \tilde{\theta}_1,\dots,\tilde{\theta}_n\rangle
  \end{equation}
The dressed form factor is, by the definition, independent of the discretization $\{\theta\}$ and simply depends on the choice of $\vartheta$. Therefore
\begin{equation}
  \langle \vartheta|\mathcal{O}|\vartheta, \theta_1, \dots, \theta_n\rangle = \exp\left( - \frac{1}{2}\delta S[\vartheta, \theta_1, \dots, \theta_n] \right)  \lim_{\rm th}\, \langle\{\tilde{\theta}\}\vert\mathcal{O}\vert\{\tilde{\theta^\prime}\}, \tilde{\theta}_1,\dots,\tilde{\theta}_n\rangle_{\rm dressed}.\label{nosoftmodes}
\end{equation}
At the present state of the development of the Bethe Ansatz techniques the evaluation of the dressed form factors is a difficult task. However certain progress in this direction was achieved in~\cite{2011_Shashi_PRB_84,1742-5468-2012-09-P09001}. Here, following the logic presented in~\cite{1742-5468-2015-2-P02019} and further developed in~\cite{SciPostPhys.1.2.015,2018JSMTE..03.3102D}, we approximate the sum over soft modes by simply picking one specific, reference state. The choice of this state is operator dependent and it also might depend on the excitations $\tilde{\theta}_1, \dots, \tilde{\theta}_n$. Therefore we replace the dressed form factor with a single form factor times the phase space of soft modes, which is $\exp(\delta S[\vartheta])$,
\begin{equation}
  \lim_{\rm th}\, \langle\{\tilde{\theta}\}\vert\mathcal{O}\vert\{\tilde{\theta^\prime}\}, \tilde{\theta}_1,\dots,\tilde{\theta}_n\rangle_{\rm dressed} \rightarrow \exp(\delta S[\vartheta]) \lim_{\rm th}\, \langle\{\tilde{\theta}\}\vert\mathcal{O}\vert\{\tilde{\theta}\}, \tilde{\theta}_1,\dots,\tilde{\theta}_n\rangle
\end{equation}
The thermodynamic form factor then becomes
\begin{equation}
  \langle \vartheta|\mathcal{O}|\vartheta, \theta_1, \dots, \theta_n\rangle = \exp\left(\frac{1}{2}\delta S[\vartheta, \theta_1, \dots, \theta_n] \right)  \lim_{\rm th}\, \langle\{\tilde{\theta}\}\vert\mathcal{O}\vert\{\tilde{\theta}\}, \tilde{\theta}_1,\dots,\tilde{\theta}_n\rangle.
\end{equation}
At this stage it is worth noting that the axioms we will derive will not really rely on approximating the dressed form factor  by a single reference state form factor. Instead the assumption is that the new factors coming from exchaning particles or from periodicity axiom are the same for the dressed form factor and for the form factor of the reference state. Finally, we write form factors as functions of quantum numbers
\beq
\langle \vartheta\vert \mathcal{O}\vert \vartheta,\{\theta_i\}\rangle = \exp\left(\frac{1}{2}\delta S[\vartheta, \{I_i\}] \right)\lim_{\rm th}  \sqrt{\rho(\{\tilde{\theta}\})\rho(\{\tilde{\theta}^\prime\},\{\tilde{\theta}_j\})}\,\langle \{I\}\vert\mathcal{O}\vert \{ I \},\{ I_i \}\rangle.\nonumber
\eeq
The formula, relating the finite volume form factors with their thermodynamic limit, is the main result of this section. In the next section we will use it to conjecture the axioms for the finite background form factors from the axioms for the vacuum ones. The derivation presented here follows closely conceptually similar derivations in non-relativistic integrable models shown for example in~\cite{1742-5468-2015-2-P02019,QA_Caux}.

Note that we assume we know, as an input, the one-point expectation value,  $\langle \vartheta\vert \mathcal{O}\vert\vartheta\rangle/\langle\vartheta\vert\vartheta\rangle$, which we can think of as a zero-particle form factor in our approach. In general, this value can be computed from the one-point Leclair-Mussardo formula \cite{LECLAIR1999624}, which is an expansion in terms of standard zero-background form factors.  For some specific operators, more convenient closed-form expressions may be found, which essentially resum the Leclair-Mussardo series. One example is when the operator is given by the trace of the energy momentum tensor, $\mathcal{O}=T^\mu_\mu$, in which case an exact thermal expectation value can be computed directly from the thermodynamic Bethe ansatz formalism \cite{ZAMOLODCHIKOV1990695}. Another more recent example is the conjectured Negro-Smirnov formula \cite{Negro:2013wga}, which gives a closed-form expression for the expectation value of vertex operators in sinh-Gordon on a given thermodynamic background.

The analysis presented so far works for fermions with antiperiodic boundary conditions. However, for bosons with periodic boundary conditions there is a small caveat. The problem is that in such
theories quantum numbers change from integers to half-odd integers when we add or remove a particle. Consider a form factor with a single particle in a bosonic theory
\begin{equation}
  \langle \vartheta|\mathcal{O}|\vartheta, \theta_1\rangle.
\end{equation}
Following the procedure outlined above we can discretize state $|\vartheta\rangle$ to find a set of integers $\{I\}$ defining a finite system reference state $|\{I\}\rangle$. However what quantum numbers $\{\tilde{I}\}$ should we choose to represent background of $|\vartheta, \theta_1\rangle$? They cannot be the same quantum numbers as $\{I\}$ because there is now one more particle. Therefore $\{\tilde{I}\}$ must differ from $\{I\}$ by $\pm 1/2$. In principle any choice of $\pm 1/2$ is valid. However, guided by the physical intuition, we choose the quantum numbers $\{\tilde{I}\}$ to spread from the newly inserted particle,
\begin{equation}
  \tilde{I}_j = \begin{cases}
    I_j - \frac{1}{2}, \quad I_j < J,\\
    I_j + \frac{1}{2}, \quad I_j > J,
    \end{cases}
\end{equation}
where $J$ is the quantum number corresponding to the extra particle with rapidity $\theta_1$ in state $|\vartheta, \theta_1\rangle$.

The shift of quantum numbers by a factor $1/2$ causes an additional change to the rapidities of the state $|\{\tilde{I}\};J\rangle$ comparing with the state $|\{I\}\rangle$. In the thermodynamic limit this leads to a modification of a back-flow function. The back-flow functions obeys the modified integral equation
\begin{equation}
  2\pi F_B(\tilde{\theta}|\theta_1) = \delta_B(\tilde{\theta} - \theta_1) + \int d\theta' \vartheta(\theta') \varphi(\tilde{\theta} - \theta')F_B(\theta'| \theta_1),\qquad  \delta_B(\theta) = \frac{1}{2}\left(1 - 2\Theta_H(\theta)\right) + \delta_0(\theta),
\end{equation}
where $\Theta_H(\theta)$ is the Heaviside function defined to be 
\begin{equation}
  \Theta_H(\theta) = \begin{cases}
    0, \quad \theta < 0, \\
    \frac{1}{2}, \quad \theta = 0, \\
    1, \quad \theta > 0.
  \end{cases}
\end{equation}
This is the back-flow function that we will use for bosonic theories.

The back-flow function for the hole is again a minus back-flow function for a particle
\begin{equation}
  2\pi F_B(\tilde{\theta}|\theta_1 + i\pi) = -2\pi F_B(\tilde{\theta}|\theta_1).
\end{equation}
In the case of a multiparticle state, the total back-flow is still a sum of the single excitation back-flows
\begin{equation}
  F_B(\tilde{\theta}|\theta_1, \dots, \theta_n) = \sum_{i=1}^n F_B(\tilde{\theta}|\theta_i).
\end{equation}
In the rest of the manuscript we adopt the following shorthand notation. Given a set of quantum numbers $\{I\}$ of the background state $|\vartheta\rangle$ we denote the finite volume counterpart of the excited state $|\vartheta, \theta_1, \dots, \theta_n\rangle$ as
\begin{equation}
  |\{\bar{I}\}, I_1, \dots, I_n\rangle.
\end{equation}
The bar reminds us that the set of quantum numbers $\{\bar{I}\}$ is shifted, according to the above prescription, due the presence of $I_1, \dots, I_n$. We will use the same notation also for theories with antiperiodic boundary conditions remembering that in that case there is no shift. Also, in what follows, we will write just $F(\theta|\theta)$ and $\delta(\theta)$ keeping in mind that both functions have to be choosen appropriately as well.

\section{Form factors on a finite energy density background}\label{sec:bootstrap_background}

We now want to find axioms for the (normalized) form factor, which we define as the function 
\beq
f_{\vartheta}^\mathcal{O}(\theta_1,\dots,\theta_n)\equiv \frac{ \langle \vartheta\vert \mathcal{O}(0)\vert\vartheta, \theta_1,\dots,\theta_n\rangle}{\langle\vartheta\vert\vartheta\rangle}=\lim_{\rm th}\sqrt{\frac{\rho(\{\tilde{\theta}^\prime\},\tilde{\theta}_1,\dots,\tilde{\theta}_n)}{\rho(\{\tilde{\theta}\})}}\langle\{I\}\vert\mathcal{O}(0)\vert \{\bar{I}\},I_1,\dots,I_n\rangle\label{normalization}.
\eeq
 There are several reasons to choose this particular normalization. First, this normalization ensures that in the low density limit where $\vartheta(\theta)\to0$, the form factor reduces to the standard form factor defined on the vacuum,
 \beq
 \lim_{\vartheta(\theta)\to0}f_\vartheta(\theta_1,\dots,\theta_n)=f(\theta_1,\dots,\theta_n).\nonumber
 \eeq
 The normalization (\ref{normalization}) also ensures that the form factor axioms we will derive in this section look very natural and similar to the axioms obeyed by the standard vacuum form factors. Finally, we will show  this normalization leads to a simple and natural expression for correlation functions.

 Our strategy is now the following. We start with axioms for  form factors with a finite set of excitations at finite volume. The finite-volume form factors can be expressed in terms of infinite volume form factors, via Eq. (\ref{finitevolumeff}). We then take the thermodynamic limit, taking the large-volume limit, but keeping the energy density constant. This leads to relations that we propose to be axioms for (thermodynamic) finite energy density form factors. 
 
    The finite size form factors we need can be written as 
\beq
\langle \{I\}\vert\mathcal{O}\vert \{\bar{I}\},I_1,\dots,I_n\rangle = \frac{f^\mathcal{O}(\{\tilde{\theta}+\pi{\rm i}\},\{\tilde{\theta}^\prime\},\tilde{\theta}_1,\dots,\tilde{\theta}_n)}{\sqrt{\rho(\{\tilde{\theta}\})\rho(\{\tilde{\theta}^\prime\},\tilde{\theta}_1,\dots,\tilde{\theta}_n)}}+O\left(e^{-\mu L}\right).\label{finitevolumeff2}
\eeq

The form factor $\langle \{I\}\vert \mathcal{O}\vert \{\bar{I}\},\{I_i\}\rangle$ contains no disconnected pieces, since the rapidities corresponding to $\{I\}$ and $\{\bar{I}\}$ in the incoming and outgoing state do not match, as the rapidities in the incoming state have been displaced by the additional particles, $\{I_i\}$.

\begin{figure}
  \center
  \includegraphics[scale=0.52]{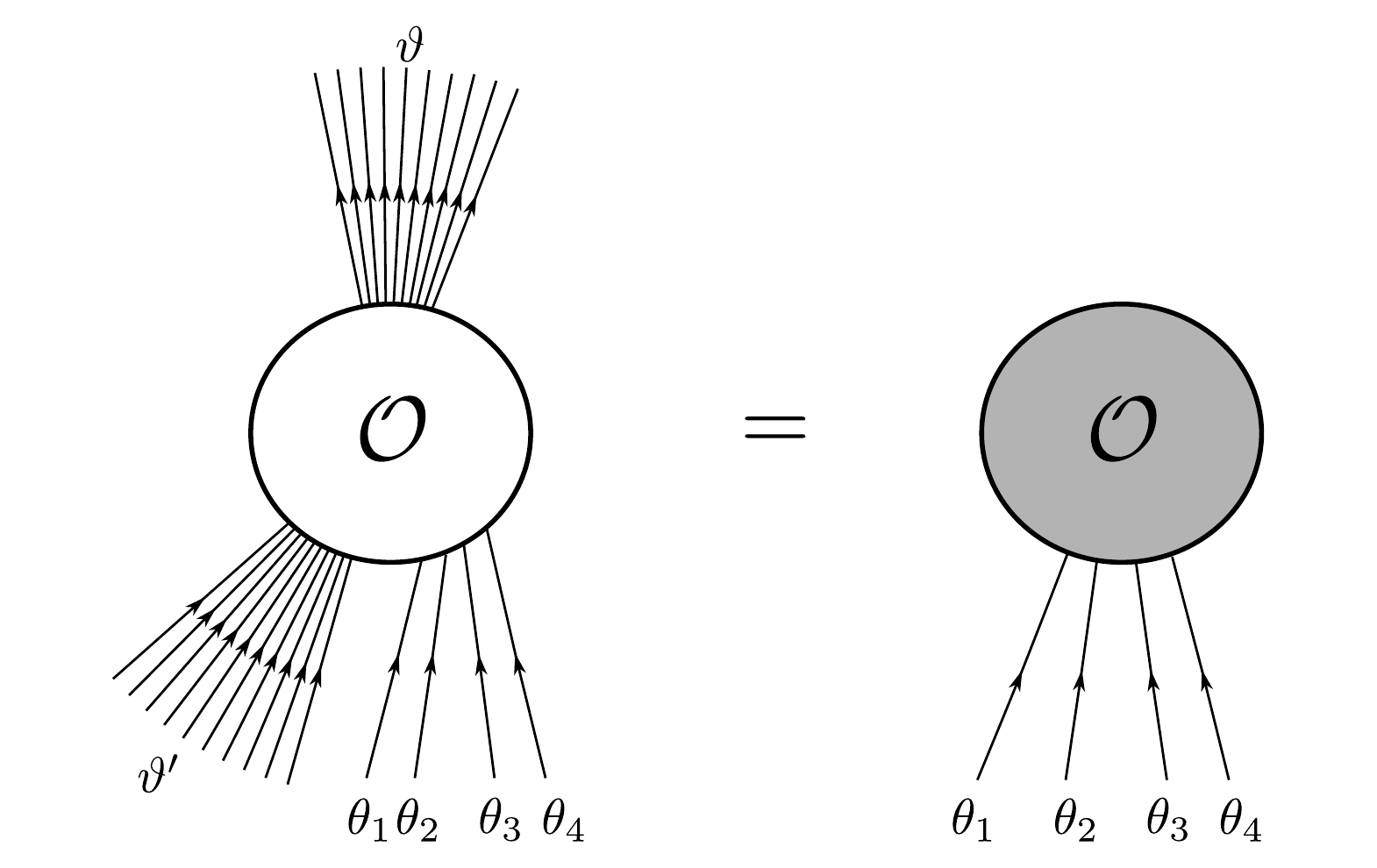}
  \caption{Graphical definition of the background-dependent form factors.}
  \label{fig:finite_bg_ff}
\end{figure}

\begin{enumerate}

\item{\it Lorentz symmetry}

It is no longer true that $f_\vartheta^\mathcal{O}(\theta_1,\dots,\theta_n)$, for a scalar operator, $\mathcal{O}$, should be invariant under a Lorentz boost of the form $\theta_i\to\theta_i+\alpha$. This is because the new rapidities can be compared with those of the particles in the background. This implies that the one-particle form factor, $f_\vartheta^\mathcal{O}(\theta_1)$, is not necessarily a constant, yet the dependence on $\theta_1$ is restricted by the periodicity axiom, as we shall see later.

\item{\it Scattering axiom}

We now investigate what happens when we exchange two adjacent particles in the form factor,
\beq
f_\vartheta^\mathcal{O}(\theta_1,\dots,\theta_i,\theta_{i+1},\dots,\theta_n)\to f_\vartheta^\mathcal{O}(\theta_1,\dots,\theta_{i+1},\theta_i,\dots,\theta_n).\nonumber
\eeq
At finite volume, the corresponding form factor is
\beq
\langle \{I\}\vert \mathcal{O}\vert \{\bar{I}\},I_1,\dots,I_i,I_{i+1},\dots,I_n\rangle=\frac{f^\mathcal{O}(\{\tilde{\theta}+\pi{\rm i}\},\{\tilde{\theta}^\prime\},\tilde{\theta}_1,\dots,\tilde{\theta}_i,\tilde{\theta}_{i+1},\dots,\tilde{\theta}_n)}{\sqrt{\rho(\{\tilde{\theta}\})\rho(\{\tilde{\theta}^\prime\},\tilde{\theta}_1,\dots,\tilde{\theta}_i,\tilde{\theta}_{i+1},\dots,\tilde{\theta}_{n})}},\label{scatteringone}
\eeq
The form factor in the numerator in the right-hand side of (\ref{scatteringone}) is the standard, infinite-volume form factor. Scattering in this form factor is a local process, between two neighboring particles, and it is not affected by the presence of background particles.  The $i$-th and $(i+1)$-th particles can be exchanged by multiplying with a factor of the S-matrix, $S(\theta_i-\theta_{i+1})$. The Jacobian in the denominator in (\ref{scatteringone}) is symmetric under exchange of two rapidities. In the finite-volume form factor, we can therefore exchange two particles as
\beq
\langle \{I\}\vert \mathcal{O}\vert \{\bar{I}\},I_1,\dots,I_i,I_{i+1},\dots,I_n\rangle=S(\tilde{\theta}_i- \tilde{\theta}_{i+1})\langle \{I\}\vert \mathcal{O}\vert \{\bar{I}\},I_1,\dots,I_{i+1},I_{i},\dots,I_n\rangle.\nonumber
\eeq

\begin{figure}
  \center
  \includegraphics[scale=0.5]{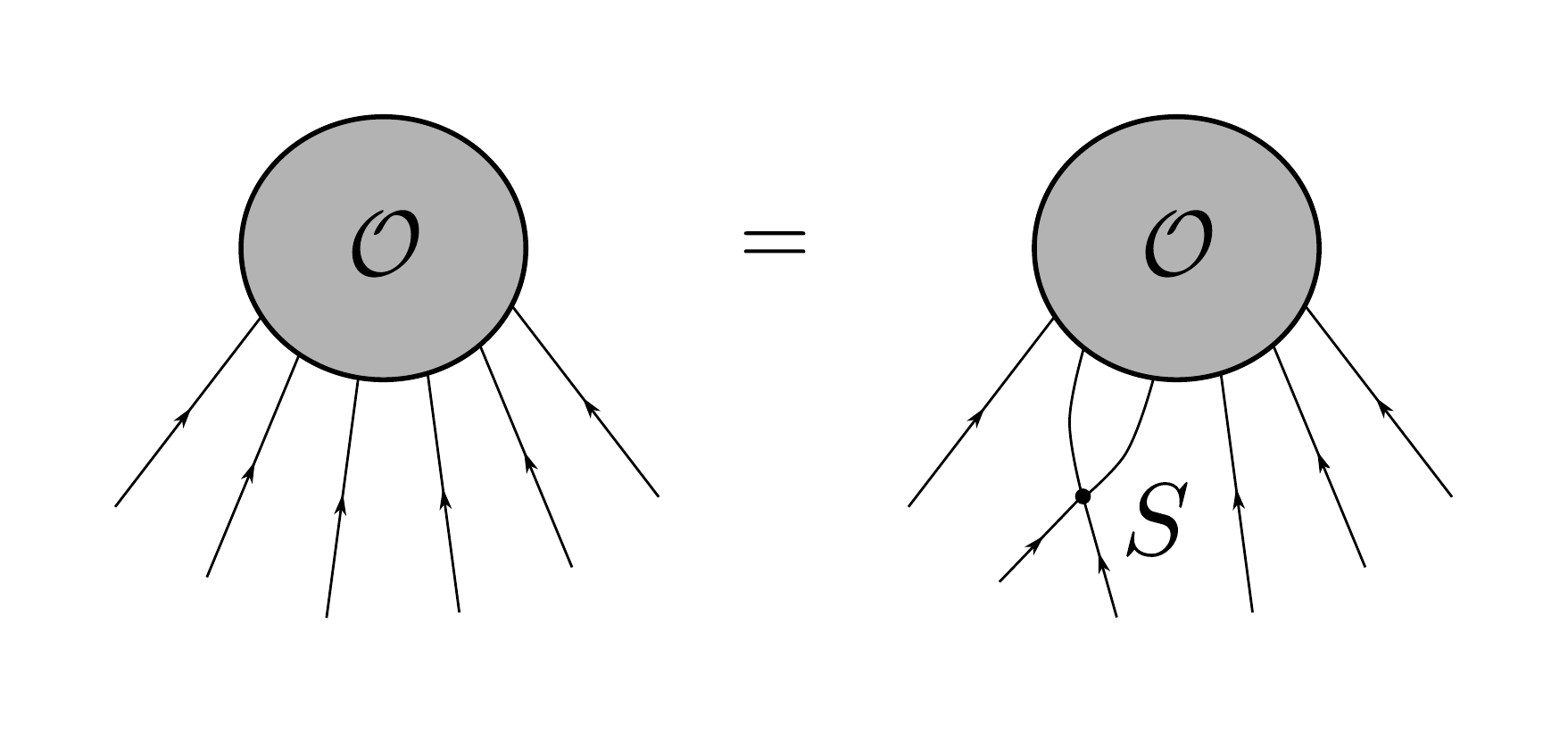}
  \caption{Scattering axiom in the presence of the background.}
  \label{fig:finite_bg_scattering}
\end{figure}

This S-matrix exchange factor is not affected by the presence of the background. As we increase the system size, and introduce more background particles accordingly, this exchange factor remains the same. In the thermodynamic limit, we can then write the scattering axiom as, see also Fig.~\ref{fig:finite_bg_scattering},
\beq
\boxed{
f_\vartheta^\mathcal{O}(\theta_1,\dots,\theta_i,\theta_{i+1},\dots,\theta_n)=S(\theta_i-\theta_{i+1}) f_\vartheta^\mathcal{O}(\theta_1,\dots,\theta_{i+1},\theta_i,\dots,\theta_n)}.\label{finitedensityscattering}
\eeq
Therefore, it takes the same form as in the vacuum

\item{\it Periodicity axiom}

\begin{figure}
  \center
  \includegraphics[scale=0.52]{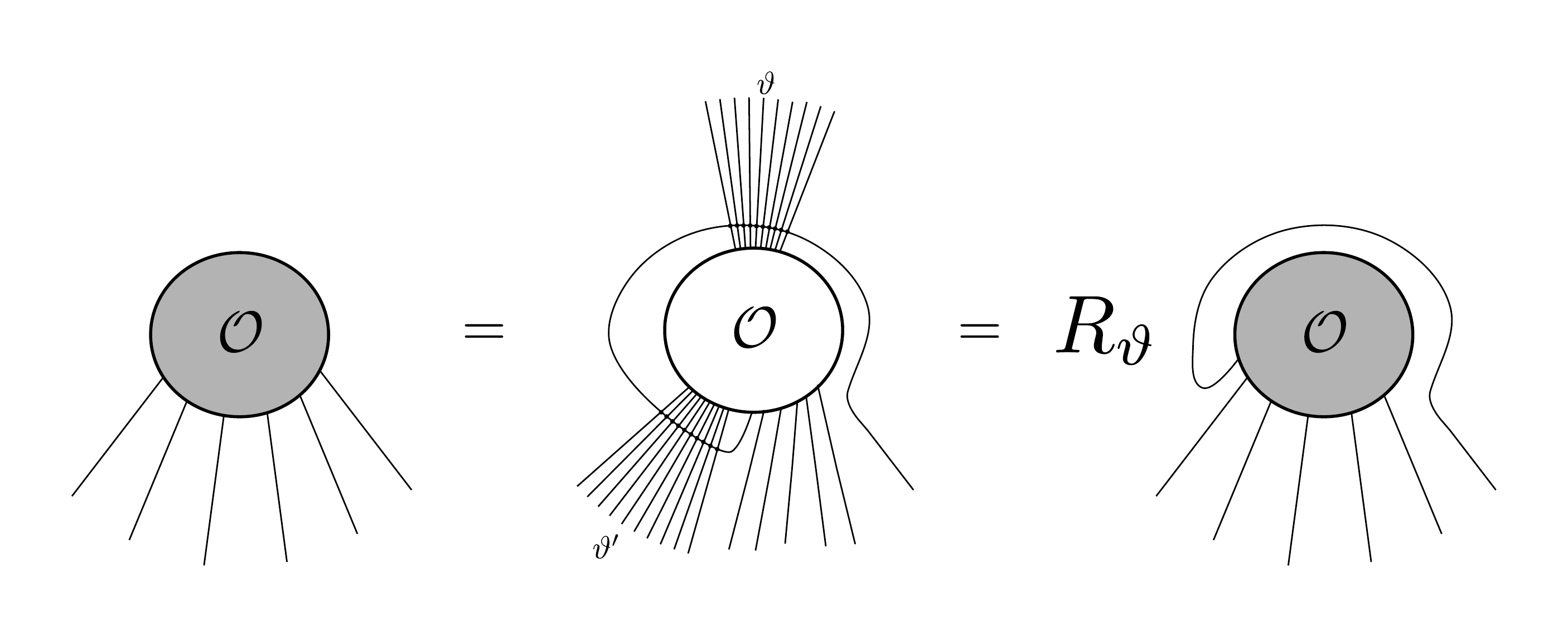}
  \caption{The periodicity axiom in the presence of the background is modified by an extra phase factor which depends on the background, the particle content and on the particle which is moved around. This extra phase factor comes from scattering of the particle going around with the background particles of the incoming and outgoing states.}\label{fig:periodicity_axiom_background}
\end{figure}

We now find what happens if we shift the last excited particle to become the first particle by applying crossing symmetry twice,
\beq
f_{\vartheta}^\mathcal{O}(\theta_1,\dots,\theta_n)\to f_\vartheta^\mathcal{O}(\theta_n+2\pi{\rm i},\theta_1,\dots,\theta_{n-1}).\nonumber
\eeq

We recall that the presence of the $n$-additional particles causes a shift in the background particles $\tilde{\theta}^\prime=\tilde{\theta}+F(\tilde{\theta}\vert\{\tilde{\theta}_n\})/{L\rho_s(\tilde{\theta})}$.  The finite-volume form factor can then be written as 
\beq
\langle \{I\}\vert \mathcal{O}\vert \{\bar{I}\},I_1,\dots,I_n\rangle=\frac{f^\mathcal{O}(\{\tilde{\theta}+\pi{\rm i}\},\{\tilde{\theta}+F(\tilde{\theta}\vert \{\tilde{\theta}_i\})/{L\rho_s(\tilde{\theta})}\},\tilde{\theta}_1,\dots,\tilde{\theta}_n)}{\sqrt{\rho(\{\tilde{\theta}\})\rho(\{\tilde{\theta}+F(\tilde{\theta}\vert \{\tilde{\theta}_i\})/{L\rho_s(\tilde{\theta})}\},\tilde{\theta}_1,\dots,\tilde{\theta}_{n})}},\nonumber
\eeq
With this expression, we are now able to apply twice crossing symmetry on the $n$-th particle, and bring it to the position before the first excited particle. After doing this process of double crossing, the $n$-th particle needs to scatter with all the particles in the background in the outgoing and incoming states, such that we can write
\beq
&&\frac{f^\mathcal{O}(\{\tilde{\theta}+\pi{\rm i}\},\{\tilde{\theta}+F(\tilde{\theta}\vert \{\tilde{\theta}_i\})/{L\rho_s(\tilde{\theta})}\},\tilde{\theta}_1,\dots,\tilde{\theta}_n)}{\sqrt{\rho(\{\tilde{\theta}\})\rho(\{\tilde{\theta}+F(\tilde{\theta}\vert \{\tilde{\theta}_i\})/{L\rho_s(\tilde{\theta})}\},\tilde{\theta}_1,\dots,\tilde{\theta}_{n})}}\nonumber\\
&&\,\,\,\,\,\,\,\,\,\,\,\,\,\,\,\,\,\,\,\,\,\,\,\,\,\,\,\,\,\,=\prod_{\{\tilde{\theta}\}} S^{-1}(\tilde{\theta}_n-\tilde{\theta})\prod_{\{\tilde{\theta}\}}S\left(\tilde{\theta}_n-\tilde{\theta}-\frac{F(\tilde{\theta}\vert\{\tilde{\theta}_i\})}{L\rho_s(\tilde{\theta})}\right)\nonumber\\
&&\,\,\,\,\,\,\,\,\,\,\,\,\,\,\,\,\,\,\,\,\,\,\,\,\,\,\,\,\,\,\,\,\,\,\,\,\,\,\,\,\times\frac{f^\mathcal{O}(\{\tilde{\theta}+\pi{\rm i}\},\{\tilde{\theta}+F(\tilde{\theta}\vert \{\tilde{\theta}_i\})/{L\rho_s(\tilde{\theta})}\},\tilde{\theta}_n+2\pi {\rm i}, \tilde{\theta}_1,\dots,\tilde{\theta}_{n-1})}{\sqrt{\rho(\{\tilde{\theta}\})\rho(\{\tilde{\theta}+F(\tilde{\theta}\vert \{\tilde{\theta}_i\})/{L\rho_s(\tilde{\theta})}\},\tilde{\theta}_n,\tilde{\theta}_1,\dots,\tilde{\theta}_{n-1})}}.\label{periodicityfinitevolume}
\eeq
We notice that the two infinite products of S-matrices in (\ref{periodicityfinitevolume}) would completely cancel each other, if the set of background rapidities in the incoming state were not shifted by $F(\tilde{\theta}\vert\{\tilde{\theta}_i\})/L\rho_s(\tilde{\theta})$. Taking the thermodynamic limit of expression (\ref{periodicityfinitevolume}), we then find the new periodicity axiom
\beq
\boxed{
f_\vartheta^\mathcal{O}(\theta_1,\dots,\theta_n)=R_\vartheta(\theta_n\vert \theta_1,\dots,\theta_n) \,f_\vartheta^\mathcal{O}(\theta_n+2\pi{\rm i},\theta_1,\dots,\theta_{n-1}),}\label{periodicitythermodynamic}
\eeq
where
\beq
R_\vartheta(\theta_n\vert \theta_1,\dots,\theta_n)=\prod_{i=1}^nR_\vartheta(\theta_n\vert\theta_i),\label{separater}
\eeq
and $R_\vartheta(\theta_n\vert\theta_j)$ is defined as 
\beq
R_\vartheta(\theta_i\vert\theta_j)=e^{i \left(2\pi F(\theta_i|\theta_j) -\delta(\theta_i-\theta_j)\right)}.\label{rdef}
\eeq
This a good moment to recall our convention according to which functions appearing in the exponent have to be choosen appropriately depending whether the theory under the consideration is fermionic or bosonic. In any case function $R_\vartheta(\theta_i\vert\theta_j)$ has the following asymptotic behavior  
\begin{equation}
\lim_{\theta_i\rightarrow \pm \infty} R_\vartheta(\theta_i\vert\theta_j)= \lim_{\theta_j\rightarrow \pm \infty} R_\vartheta(\theta_i\vert\theta_j) = 1.
\end{equation}

\item{\it Annihilation-pole axiom}\footnote{We would like to acknowledge discussions with Jacopo De Nardis that helped us in formulating this axiom.}
In the presence of the finite-density background, two excitations in the form factor may still annihilate with each other, requiring an annihilation pole, but the residue of the form factor at this pole will now involve the scattering terms with the particles on the background. Taking the thermodynamic limit of the finite-volume expression, as was done for the previous axioms, it is easy to find that the annihilation pole axiom now becomes

\begin{empheq}[box=\fbox]{align*}
&-{\rm i}{\rm Res}_{\theta_1=\theta_2}f_\vartheta^\mathcal{O}(\theta_1+\pi {\rm i},\theta_2,\theta_3,\dots,\theta_n)\nonumber\\
&\,\,\,\,\,\,\,\,\,\,\,\,\,\,=\left[1-{R}_\vartheta(\theta_2\vert\theta_3,\dots,\theta_n) S(\theta_{2}-\theta_3)\times\dots\times S(\theta_{2}-\theta_{n})\right]f_\vartheta^\mathcal{O}(\theta_3,\dots,\theta_n)
\label{dressedannihilation}
\end{empheq}
This axiom provides a recursive relation, connecting the $n$-particle and $n-2$ particle form factors.

\item{\it Cluster properties}

Again starting with the finite-volume expression (\ref{finitevolumeff}), we can study the cluster properties that follow from shifting the rapidities of some subset of particles, by the same amount, $\alpha\to\infty$. 

It is important to point out that the two-particle S-matrix typically satisfies the condition 
\beq
\lim_{\vert\theta\vert\to\infty} S(\theta)=1,\,\,\,\,\,\,\,\,\lim_{\vert \theta\vert\to\infty}\delta(\theta)=0.\nonumber
\eeq
This implies that if in the expression (\ref{finitevolumeff}), we boost the rapidities of a set of the particles by a large amount, $\alpha\to\infty$,  the corresponding Jacobian for this state is a block-diagonal matrix, where the boosted particles decouple from the rest of the particles. Using then the property that the determinant of a block-diagonal matrix equals the product of the determinants of the blocks, we find
\beq
\lim_{\alpha\to\infty}\rho(\{\tilde{\theta^\prime}\},\tilde{\theta_1},\dots,\tilde{\theta}_i,\tilde{\theta}_{i+1}-\alpha, \dots,\tilde{\theta}_n-\alpha)=\rho(\{\tilde{\theta^\prime}\},\tilde{\theta_1},\dots,\tilde{\theta}_{i})\rho(\tilde{\theta}_{i+1},\dots,\tilde{\theta}_n).\label{clusterrho}
\eeq

It can now be easily derived that \beq
&&\lim_{\alpha\to\infty}\frac{f^{\mathcal{O}_1}(\{\tilde{\theta}+\pi{\rm i}\},\{\tilde{\theta}^\prime\},\tilde{\theta}_1,\dots,\tilde{\theta}_i,\tilde{\theta}_{i+1}-\alpha,\dots,\tilde{\theta}_{n}-\alpha)}{\sqrt{\rho(\{\tilde{\theta}\})\rho(\{\tilde{\theta^\prime}\},\tilde{\theta_1},\dots,\tilde{\theta}_i,\tilde{\theta}_{i+1}-\alpha, \dots,\tilde{\theta}_n-\alpha)}}\nonumber\\
&&\nonumber\\
&&\,\,\,\,\,\,\,\,\,\,\,\,\,\,\,\,\,\,\,\,\,\,\,\,\,\,\,\,\,\,\,\,\,\,\,=\mathcal{N}_{\rm cluster}^{\mathcal{O}_1}\,\frac{f^{\mathcal{O}_2}(\{\tilde{\theta}+\pi{\rm i}\},\{\tilde{\theta}^\prime\},\tilde{\theta}_1,\dots,\tilde{\theta}_i)}{\sqrt{\rho(\{\tilde{\theta}\})\rho(\{\tilde{\theta}^\prime\},\tilde{\theta}_1,\dots,\tilde{\theta}_i)}}\times\frac{f^{\mathcal{O}_3}(\tilde{\theta}_{i+1},\dots,\tilde{\theta}_{n})}{\sqrt{\rho(\tilde{\theta}_{i+1},\dots,\tilde{\theta}_n)}}.\nonumber
\eeq

Now taking the thermodynamic limit, we  find
\beq
\boxed{
\lim_{\alpha\to\infty}f_\vartheta^{\mathcal{O}_1}(\theta_1,\dots,\theta_i,\theta_{i+1}-\alpha,\dots,\theta_n-\alpha)=\mathcal{N}_{\rm cluster}^{\mathcal{O}_1}\,f_\vartheta^{\mathcal{O}_2}(\theta_1,\dots,\theta_i)f^{\mathcal{O}_3}(\theta_{i+1},\dots,\theta_n)},\label{cluster}
\eeq
where we point out that the second form factor on the right hand side of (\ref{cluster}) is the standard, zero-background form factor.

\item{\it Bounds on growth}

The same bounds apply on the form factors, as in the zero-background case. This can be seen from the  finite volume expression for the form factors (\ref{finitevolumeff}).
The form factor on the numerator  of the right-hand-side of (\ref{finitevolumeff}) has known bounds as we take the the rapidities of any of the excitations on top of the background to be large. It is clear from the properties of the determinant of the Jacobian, (\ref{clusterrho}), that if one of the rapidities is taken to be very large, its contribution to this determinant simply factorizes. The behavior of the  denominator  then does not affect this bound.

The same bound then holds for the form factor even in the presence of the background,
\beq
\boxed{
\lim_{\vert\theta_i\vert\to\infty}f_\vartheta^\mathcal{O}(\theta_1,\dots,\theta_n)\sim e^{y_\mathcal{O}\vert \theta_i\vert}}.\label{boundongrowth}
\eeq

\item{\it Translation}

To obtain the form factors of the operator at some space-time point, $x,t$, we can use Lorentz symmetry to write
\beq
f_\vartheta^\mathcal{O}(x,t)(\theta_1,\dots,\theta_n)=e^{{\rm i}x\left[P(\vartheta;\theta_1,\dots,\theta_n)-P(\vartheta)\right]-{\rm i}t\left[E(\vartheta;\theta_1,\dots,\theta_n)-E(\vartheta)\right]}f_\vartheta^\mathcal{O}(\theta_1,\dots,\theta_n),\nonumber
\eeq
where $E(\vartheta),\,P(\vartheta)$ are the energy and momentum of the background state, respectively, and $E(\vartheta;\theta_1,\dots,\theta_n)$ $\,P(\vartheta;\theta_1,\dots,\theta_n)$ are the energy and momentum of the background state with the particle excitations on top of it, respectively.

These total energies and momenta can be expressed in terms of the dressed energy and momentum of particles, which can be obtained from the TBA formalism \cite{ZAMOLODCHIKOV1990695}, such that,
\beq
\boxed{
f_\vartheta^\mathcal{O}(x,t)(\theta_1,\dots,\theta_n)=e^{{\rm i}x\sum_i\varepsilon(\theta_i)-{\rm i}t\sum_ik(\theta_i)}f_\vartheta^\mathcal{O}(\theta_1,\dots,\theta_n)},\label{translation}
\eeq

where the dressed energy and momentum are given by (see Appendix~\ref{app:finite_volume}), respectively,
\begin{align}
  k(\theta) &= m\sinh\theta - \int d\theta' \vartheta(\theta') \cosh \theta' F(\theta'|\theta),\\
  \epsilon(\theta) &= m\cosh\theta -  \int d\theta' \vartheta(\theta') \sinh \theta' F(\theta'|\theta).
\end{align}

\end{enumerate}

\section{Analyzing the consequences of the form factors axioms}\label{sec:consequences}

We will now explore the restrictions that the new form factor axioms place on the functions $f_\vartheta(\theta_1,\dots,\theta_n)$. We point out that the procedure outlined here yields the {\it minimal} solution to all the form factor axioms. Without any further evidence to the contrary, we assume the form factors are described by the simplest solutions which satisfy all the axioms. For instance, we assume the form factors have maximally analytic structure, with no zeros or poles which are not kinematically or otherwise required.

\subsection{One-particle form factors}

We begin by examining the one-particle form factor, $f_\vartheta(\theta)$, which as we have established, now may depend on the particle's rapidity, in contrast to the zero-background case where the one-particle form factor is a constant. 

The periodicity axiom for this form factor gives the condition
\beq
f_\vartheta^\mathcal{O}(\theta)=R_\vartheta(\theta\vert \theta) f_\vartheta^\mathcal{O}(\theta+2\pi {\rm i}).\label{oneparticleperiodicity}
\eeq
We can search for a minimal solution to this equation, which we denote as the function $f_\vartheta^{\rm min}(\theta)$, which satisfies the condition while being maximally analytic. We notice that Equation (\ref{oneparticleperiodicity}) is very similar to the monodromy equation (\ref{monodromy}) satisfied by the minimal two-particle form factors at zero density background. We can therefore easily find the minimal solution to (\ref{oneparticleperiodicity}) in the same way. If we can find an expression
\beq
R_\vartheta(\theta\vert\theta)=\exp\left[\frac{1}{2}\int_{-\infty}^\infty\frac{dt}{t}C_\vartheta(t) \exp\frac{t\theta}{\rm i\pi}\right],\label{definitionc}
\eeq
then the minimal solution to (\ref{oneparticleperiodicity}) is
\beq
f_\vartheta^{\rm min}(\theta)=\exp\left[-\frac{1}{4}\int_{-\infty}^\infty\frac{dt}{t}\frac{C_\vartheta(t)}{\sinh t}\exp\left(\frac{t(\theta - i\pi)}{i\pi}\right)\right].\nonumber
\eeq
Function $R_{\vartheta}(\theta|\theta')$ is a pure phase and therefore
\beq
C_{\vartheta}^*(t) = C_{\vartheta}(-t). 
\eeq

From eq.~\eqref{definitionc} and definition~\eqref{rdef} of $R_{\vartheta}(\theta\vert\theta)$ it follows that $C_{\vartheta}(\theta)$ is related to the Fourier transform of the back-flow function $F(\theta|\theta)$. This implies that representation~\eqref{definitionc} is indeed possible. 

The general one-particle form factor can then be expressed as 
\beq
f_\vartheta^\mathcal{O}(\theta)=K_\vartheta^\mathcal{O}(\theta)f_\vartheta^{\rm min}(\theta),\nonumber
\eeq
where $K_\vartheta^\mathcal{O}(\theta)$ is a periodic function, invariant under $\theta\to\theta+2\pi{\rm i}$, which may depend on the properties of the operator, and should not have any poles (since the one particle state should not have any kinematical annihilation or bound state pole properties). These properties imply that we can express this function, in general, as the polynomial expansion
\beq
K_\vartheta^\mathcal{O}(\theta)=\sum_{n=0}^{N_\vartheta^\mathcal{O}} a_{\vartheta\,n}^\mathcal{O}\cosh^n\theta,\label{coshpolynomial}
\eeq
where all of the information about the background, $\vartheta$, and the operator, $\mathcal{O}$, is contained in the coefficients, $a_{\vartheta\, n}^\mathcal{O}$, and the series is truncated at some operator and background dependent level, $N_\vartheta^\mathcal{O}$.

From the cluster properties of the dressed form factors (\ref{cluster}), we expect the one-particle form factor to satisfy
\beq
\lim_{\theta\to-\infty} f_\vartheta^{\mathcal{O}_1}(\theta)=\mathcal{N}_{\rm cluster}^{\mathcal{O}_1}\,\frac{\langle \vartheta\vert \mathcal{O}_2\vert\vartheta\rangle}{\langle\vartheta\vert\vartheta\rangle}f^{\mathcal{O}_3}(\theta)\equiv \mathcal{N}_{\rm cluster}^{\mathcal{O}_1}\frac{\langle\vartheta\vert\mathcal{O}_2\vert\vartheta\rangle}{\langle\vartheta\vert\vartheta\rangle}f^{\mathcal{O}_3}_1, \label{limitoneparticle}
\eeq
where  the right-hand-side involves the zero-background one-particle form factor, which is a constant and does not depend on the particle's rapidity, and the finite-background expectation value, which could be determined using the one-point Leclair-Mussardo formula.   This provides a normalization for dressed form factors in terms of the zero-background form factors. For self clustering operators, such that $\mathcal{O}_1=\mathcal{O}_2=\mathcal{O}_3=\mathcal{O}$, the cluster properties reduce to,
\beq
\lim_{\theta\to-\infty} f_\vartheta^{\mathcal{O}}(\theta)=\frac{1}{\langle 0\vert\mathcal{O}\vert 0\rangle}\frac{\langle\vartheta\vert\mathcal{O}\vert\vartheta\rangle}{\langle\vartheta\vert\vartheta\rangle}f_1^\mathcal{O}.\label{oneparticleselfcluster}
\eeq

The condition (\ref{limitoneparticle}) provides an upper bound for the level, $N_\vartheta^\mathcal{O}$, of the expansion (\ref{coshpolynomial}).  The level $N_\vartheta^\mathcal{O}$ is constrained by the fact that the function $f_\vartheta^\mathcal{O}(\theta)$ must go to a constant value at large rapidities, while the powers of hyperbolic cosine grow as $\lim_{\theta\to\infty}\cosh^n\theta\to e^{n\theta}/2$. If the  minimal function $f_\vartheta^{\rm min}(\theta)$ does not exponentially decay at large rapidities, then only the constant term in (\ref{coshpolynomial}) will survive (and the constant $a_{\vartheta\,0}^\mathcal{O}$ can be fixed using  (\ref{limitoneparticle})), or $N_\vartheta^\mathcal{O}=0$.

\subsection{Form factors with two or more particles}

We can also attempt to find a minimal solution for the two-particle form factor by imposing the restrictions from the scattering and periodicity axioms. Combining these two axioms, we find a modified monodromy equation
\beq
R_\vartheta(\theta_1\vert\theta_1,\theta_2)S(\theta_1-\theta_2)f_\vartheta^\mathcal{O}(\theta_2,\theta_1)=f_\vartheta^\mathcal{O}(\theta_2,\theta_1-2\pi{\rm i}),\label{dressedmonodromy}
\eeq
 One can similarly derive a monodromy equation by applying the periodicity axiom on  the $\theta_2$ particle instead of the $\theta_1$ particle. 

A minimal solution for the two-particle form factor satisfying the monodromy equation can be found of the form
\beq
f_\vartheta^{\rm min}(\theta_1,\theta_2)=A_\vartheta^{\rm min}(\theta_1;\theta_2)A_\vartheta^{\rm min}(\theta_2;\theta_1) f^{\rm min}(\theta_1-\theta_2),\nonumber
\eeq
where $f^{\rm min}(\theta)$ is the standard (background independent) minimal form factor (\ref{fmin}). We define the function $A_\vartheta^{\rm min}(\theta;\theta_1,\dots,\theta_n)$ to satisfy the monodromy conditions
\beq
A_\vartheta^{\rm min}(\theta;\theta_1,\dots,\theta_i+2\pi {\rm i},\dots,\theta_n)=A_\vartheta^{\rm min}(\theta;\theta_1,\dots,\theta_i,\dots,\theta_n),\label{monodromyaone}
\eeq
for $i\in [1,n]$, and
\beq
R_\vartheta(\theta\vert\theta)R(\theta\vert\theta_1\dots\theta_n) A_\vartheta^{\rm min}(\theta+2\pi 
{\rm i};\theta_1,\dots,\theta_n)=A_\vartheta^{\rm min}(\theta;\theta_1,\dots,\theta_n).\label{monodromyatwo}
\eeq
A minimal solution to the monodromy equations (\ref{monodromyaone}) and (\ref{monodromyatwo}) can again be found in terms of the function $C_\vartheta(t)$ defined in (\ref{definitionc}), and a function $\tilde{C}_\vartheta(t\vert \theta_1,\dots,\theta_n)$ defined by
\beq
R(\theta\vert \theta_1,\dots,\theta_n)=\exp\left[\frac{1}{2}\int_{-\infty}^\infty\frac{dt}{t} \tilde{C}_\vartheta(t\vert \theta_1,\dots,\theta_n) \exp\frac{t\theta}{\rm i\pi}\right],\label{definitiontildec}
\eeq
The minimal solution of (\ref{monodromyaone}) and (\ref{monodromyatwo}) is then given by
\beq
A_\vartheta^{\rm min}(\theta;\theta_1,\dots,\theta_n)=\exp\left[-\frac{1}{4}\int_{-\infty}^\infty\frac{dt}{t}\frac{C_\vartheta(t)+\tilde{C}_\vartheta(t\vert \theta_1,\dots,\theta_n)}{\sinh t}\exp\left(\frac{t(\theta - i\pi)}{i\pi}\right)\right]\label{minimalageneral}
\eeq
Again, representation \eqref{definitiontildec} relates $\tilde{C}_\vartheta(t\vert\theta_1, \dots, \theta_n)$ to the Fourier transform of  $F(\theta\vert\theta_1,\dots, \theta_n)$. Moreover, the linearity of the back-flow functions in excitations implies 
\beq
\tilde{C}_{\vartheta}(t\vert\theta_1, \dots, \theta_n) = \sum_{j=1}^n \tilde{C}_{\vartheta}(t|\theta_j).
\eeq

This procedure to find minimal solutions can be easily generalized to $n$-particle form factors, which can then be generally expressed as
\beq
\boxed{
f_\vartheta^\mathcal{O}(\theta_1,\dots,\theta_n)=K_\vartheta^\mathcal{O}(\theta_1,\dots,\theta_n)\prod_i A_\vartheta^{\rm min}(\theta_i;\theta_1,\dots,\theta_{i-1},\theta_{i+1},\dots,\theta_n)\prod_{i<j}f^{\rm min}(\theta_i-\theta_j)},\label{generalformfactorwithdensity}
\eeq
where $K_\vartheta^\mathcal{O}(\theta_1,\dots,\theta_n)$ is a periodic function in all the rapidities, symmetric under exchange of two rapidities, and contains all the information about the necessary annihilation and bound state poles, and any information identifying the operator, $\mathcal{O}$.

Functions $K_\vartheta^\mathcal{O}(\theta_1,\dots,\theta_n)$ can be furher restricted. 
First of all, the annihilation pole axiom can be applied to derive recursive relations between the functions $K_\vartheta(\theta_1,\dots,\theta_n)$ and $K_\vartheta(\theta_3,\dots,\theta_n)$. Second, it is worth noting, that the location of annihilation poles is at the same values of rapidities as in the zero-background form factors. We can therefore assume the $n$-particle form factors are given by the expression (\ref{generalformfactorwithdensity}), with the function
\beq
K_\vartheta^\mathcal{O}=\frac{Q_\vartheta^\mathcal{O}(\theta_1,\dots,\theta_n)}{D(\theta_1,\dots,\theta_n)},\nonumber
\eeq
where the denominator function contains all the necessary pole structure, and is not modified by the background density. The dependence on the background and on the operator is only in the numerator, $Q_\vartheta^\mathcal{O}(\theta_1,\dots,\theta_n)$. Unlike the zero-background case, the numerator $Q_\vartheta^\mathcal{O}(\theta_1,\dots,\theta_n)$ is not necessarily a function of only rapidity diferences, $\theta_{ij}$, but it can depend on each rapidity individually.

With this  we conclude our general analysis of the form factors axioms. In section~\ref{sec:SinhGordon} we will compute the minimal form factors in the Sinh-Gordon model. We will also determine full form factors for the special class of the vertex operators. Before doing so, in the following two sections we consider the structure of the correlation functions in a state with a finite energy density.

\section{Correlation function}\label{sec:corr_func}

In this section we find an expression for the two-point correlation function, in the presence of a finite density background
\beq
C_\vartheta^{\mathcal{O},\mathcal{O}^\prime}(x,t)\equiv\frac{\langle \vartheta\vert \mathcal{O}(x,t)\mathcal{O}^\prime(0,0)\vert\vartheta\rangle}{\langle\vartheta\vert\vartheta\rangle} .\label{definitioncorrelator}
\eeq
Such correlation functions, evaluated on finite-energy-density states have many applications in strongly correlated systems. For instance, by properly choosing the function $\vartheta(\theta)$, these correlators coincide in the thermodynamic limit with finite temperature \cite{ZAMOLODCHIKOV1990695}, or generalized-Gibbs-ensemble averages \cite{Fioretto:2009yq}. The procedure, that we follow, is to write down first a correlation function in the infinite volume over a vaccum. Then we transform the formula to the finite volume. Finally we take the thermodynamic limit keeping the density of particles fixed.

At infinite volume, the vacuum correlation function can be computed by inserting a complete set of intermediate states between operators
\begin{align}
  \langle 0 \vert \mathcal{O}(x,t)\mathcal{O}^\prime(0,0)\vert 0\rangle =& \sum_{n=0}^{\infty} \left(\prod_{k=1}^n \int_{-\infty}^{\infty} \frac{d\theta}{2\pi}\right) \frac{1}{n!}f^{\mathcal{O}}(\theta_1, \dots, \theta_n)\left( f^{\mathcal{O}'}(\theta_1, \dots, \theta_n) \right)^* \nonumber \\
  &\times  \exp\left({\rm i}x m\sum_{k=1}^n \sinh\theta_k - {\rm i}tm \sum_{k=1}^n \cosh\theta_k \right).
\end{align}
In  finite volume, rapidities are quantized and are intertwined with each other as solutions of the Bethe equations. Therefore to span the Hilbert space it is more convenient to use integer quantum numbers. The correlation functions is
\begin{align}
  \langle 0 \vert \mathcal{O}(x,t)\mathcal{O}^\prime(0,0)\vert 0\rangle_L =& \sum_{n=0}^{\infty} \sum_{\{I\}} \frac{1}{n!}\langle 0\vert\mathcal{O}\vert I_1, \dots, I_n\rangle \langle I_n, \dots, I_1\vert \mathcal{O}'\vert 0\rangle \nonumber \\
  &\times  \exp\left({\rm i}x m\sum_{k=1}^n \sinh\tilde{\theta}_k - {\rm i}tm \sum_{k=1}^n \cosh\tilde{\theta}_k \right).
\end{align}
It is straightforward to promote this vacuum correlation function to a correlation function over an arbitrary finite volume state specified by a set of quantum numbers $\{J\}$
\begin{align}
  \langle \{J\} \vert \mathcal{O}(x,t)\mathcal{O}^\prime(0,0)\vert \{J\}\rangle_L =& \exp(-{\rm i}xP_J + {\rm i}t E_J)\sum_{n=0}^{\infty}\frac{1}{n!} \sum_{\{I\}} \langle \{J\} \vert\mathcal{O}\vert I_1, \dots, I_n\rangle \langle I_n, \dots, I_1\vert \mathcal{O}'\vert \{J\}\rangle \nonumber \\
  &\times  \exp\left({\rm i}x m\sum_{k=1}^n \sinh\tilde{\theta}_k - {\rm i}tm \sum_{k=1}^n \cosh\tilde{\theta}_k \right),
\end{align}
where $E_J$ and $P_J$ are the total energy and momentum, respectively, of the background set of particles $\{J\}$. 

We are interested now in situations in which the state $\{J\}$ in the thermodynamic limit follows the particle occupation number $\vartheta(\theta)$ and define
\begin{align}
 C_\vartheta^{\mathcal{O},\mathcal{O}^\prime}(x,t) = \lim_{\rm th} \langle \{J\} \vert \mathcal{O}(x,t)\mathcal{O}^\prime(0,0)\vert \{J\}\rangle_L. 
\end{align}
To consider the thermodynamic limit of the correlation function, it is useful to change the point of view on the summation over the intermediate states. Instead of summing over quantum numbers, we should sum over quantum number representing differences from the state $\{J\}$. In this way the form factors are evaluated on states which are similar to each other. This change is motivated by the assumption that local operators connect only states which are macroscopically the same. To this end we redefine the form factors
\begin{align}
  \langle \{J\} \vert\mathcal{O}\vert I_1, \dots, I_n\rangle  \rightarrow \langle \{J\} \vert\mathcal{O}\vert \{\bar{J}\},  I_1^{\sigma_1}, \dots, I_m^{\sigma_m}\rangle .
\end{align}
Quantum numbers  $\{I_1^{\sigma_1}, \dots, I_m^{\sigma_m}\}$ specify the excitations over the background $\{\bar{J}\}$. \footnote{Recall that in our notation quantum numbers $\{\bar{J}\}$ are shfited, for bosonic theories with periodic boundary conditions, with respect to $\{J\}$, and that the shift depends on the position of excitations $I_1^{\sigma_1}, \dots, I_m^{\sigma_m}$, as explained at the end of Section~\ref{sec:QFT_finite_volume}.} The additional labels on them, $\sigma_i=\pm 1$, distinguish particles from holes. The positive sign stands for the introducion of a particle and a new quantum number, $I_i$, which is not included in the set of background numbers, $\{J\}$. The negative sign, $\sigma_i=-1$ stands for the introduction of a ``hole", that is, removing a particle with quantum number $I_i$, from the background, $\{J\}$. As mentioned above, for reasonable local physical operators we can assume that in the thermodynamic limit the matrix operators $\langle \{J\} \vert\mathcal{O}\vert \{\bar{J}\},  I_1^{\sigma_1}, \dots, I_n^{\sigma_n}\rangle$ are vanishing for large number of excitations $n$. The correlation function becomes
\begin{align}
\langle \{J\} \vert \mathcal{O}(x,t)\mathcal{O}^\prime(0,0)\vert \{\bar{J}\}\rangle_L =& \sum_{n=0}^{\infty} \sum_{\{I^\sigma\}} \frac{1}{n!}\langle \{J\} \vert\mathcal{O}\vert \{\bar{J}\}, I_1^{\sigma_1}, \dots, I_n^{\sigma_n}\rangle \langle  I_n^{\sigma_n}, \dots, I_1^{\sigma_1},\{\bar{J}\}\vert \mathcal{O}'\vert \{J\}\rangle \nonumber \\
  &\times \exp\left[-{\rm i}x(P_J-P^\prime_J) +{\rm  i}t (E_J-E^\prime_J)\right] \nonumber\\
  &\times\exp\left({\rm i}x m\sum_{k=1}^n \sigma_k\sinh\theta_k - {\rm i}tm \sum_{k=1}^n\sigma_k \cosh\theta_k \right),\label{prethermodynamic}
\end{align}
where $E^\prime_J$ and $P^\prime_J$ are the total energy and momentum, respectively, of the particles $\{J\}$, in the presence of the additional $n$ particles and holes. The sum $\sum_{\{I^\sigma\}}$ is over available integers, that is, for particles, $\sigma_k=+1$ we sum only over integer values which are not included in the set, $\{J\}$, and for holes, $\sigma_k=-1$, we sum only over the values of integers included in $\{J\}$.

We want now to take the thermodynamic limit of the correlation function. In that case, the background energy and momentum differences, $E_J-E_J^\prime$ and $P_J-P_J^\prime$ can be absorbed into the energy and momentum contributions of the excitations, by substituting with their dressed energy and momentum, $\varepsilon(\theta)$ and $k(\theta)$, respectively. Furthermore, the sum over integers, $\{I^\sigma\}$ can be replaced by a smooth integral over rapidities.  This can be explicitly done using the technique introduced in \cite{Pozsgay:2010cr}, where the sum over some function $f(\tilde{\theta}_k)$ is expressed as
\beq
\frac{1}{L}\sum_{I_k^{\sigma_k}}f(\tilde{\theta}_k)=\sum_{I_k}\oint_{I_k}\frac{dz}{2\pi}\frac{ f(z)\rho_{\sigma_k}(z)}{e^{{\rm i}LQ(z)}-1}\label{contours},
\eeq
where the contour of the integrals are around the possible integer solutions of the Bethe equations, where there is a simple pole in the denominator (these poles lie on the real axis of $\theta_k$), and $\rho_{\sigma_k}$ is a distribution ensuring that we only sum over available integer values for the given excitation. In the thermodynamic limit, these become $\rho_{-1}(\theta)=\rho_p(\theta)$, and $\rho_{+1}(\theta)=\rho_h(\theta)$.  We have also defined the shorthand notation $Q(\theta_k)=Q_k(\vartheta,\theta_1,\dots,\theta_n)$. 

Focusing on one type of excitations, particles or holes, the sum over the different contours can be replaced by a single contour encircling the real line, such that
\beq
\frac{1}{L}\sum_{I_k^{\sigma}}f(\tilde{\theta}_k)=\left(\int_{\mathbb{R}+{\rm i}\epsilon}-\int_{\mathbb{R}-{\rm i}\epsilon}\right)\frac{dz}{2\pi}\frac{f(z)\rho_{\sigma}(z)}{e^{{\rm i}LQ(z)}-1} -{\rm i} \sum_{\bar{z} \in {\rm poles}(f(z),\Gamma^\epsilon)}{\rm res}\left(\frac{f(z)\rho_{\sigma}(z)}{e^{{\rm i}LQ(z)}-1}, \bar{z}\right),
\eeq
where we introduced
\begin{equation}
  {\rm res}(f(z), \bar{z}) = \frac{1}{(n-1)!} \lim_{z\rightarrow \bar{z}} \frac{d^{n-1}}{dz^{n-1}}\left((z-\bar{z})^n f(z)\right), 
\end{equation}
for $n$ being the order of the pole of $f(z)$ at $\bar{z}$. 
where the second term, and the fact that rapidity integrals are slightly shifted off the real axis, ensure the removal of any poles in the neighborhood of the real line ($\Gamma^\epsilon$), which are not the poles corresponding to the integer solutions, $I_k^{\sigma}$. In our case,  the function $f(z)$ has such poles, which are interpreted as annihilation poles of the form factors, every time the rapidity of a given particle and a hole coincide. Since  the terms in the correlation function involve the form factors {\it squared}, these will contribute double poles in $\Gamma^\epsilon$. In the thermodynamic limit,  the integration below the real axis, $\mathbb{R}-{\rm i}\epsilon$ vanishes, such that
\beq
\lim_{\rm th}\frac{1}{L}\sum_{I_k^{\sigma}}f(\tilde{\theta}_k)=\lim_{\epsilon\to0}\int_{-\infty+{\rm i}\epsilon}^{\infty+{\rm i}\epsilon}f(z)\rho_{\sigma}(z)dz - i\sum_{\bar{z} \in {\rm poles}(f(z),\Gamma^\epsilon)}{\rm res}\left(\frac{f(z)\rho_{\sigma}(z)}{e^{{\rm i}LQ(z)}-1}, \bar{z}\right)\label{finitepart}
\eeq
We point out that the expression (\ref{finitepart}) is finite, and the possibly divergent contributions from the (double) annihilation poles are systematically regularized. It may still be a challenging task in practice to compute explicitly the contributions from the residues in the right-hand side of (\ref{finitepart}) (see similar calculations for the few-particle contributions in \cite{Pozsgay:2010cr}), but the prescription is nonetheless finite and unambiguous. We can {\it define} the integration symbol, $\fint$, to denote this precise regularization
\beq
\fint_{-\infty}^\infty f(z)\rho_{\sigma}(z)dz\equiv \lim_{\rm th}\frac{1}{L}\sum_{I_k^{\sigma}}f(\tilde{\theta}_k),
\eeq
where the integration is shifted from the real axis by ${\rm i}\epsilon$ and the residues from the double poles are subtracted as prescribed.

We can now take the thermodynamic limit of (\ref{prethermodynamic}) to arrive at our final expression for the correlation function
\begin{empheq}[box=\fbox]{align*}
 C_\vartheta^{\mathcal{O},\mathcal{O}^\prime}(x,t)=& \sum_{n=0} \sum_{\sigma_i = \pm 1}\left(\prod_{k=1}^n \fint_{-\infty}^{\infty} \frac{d\theta}{2\pi}\vartheta_{\sigma_k}(\theta_k) \right) f_{\vartheta}^{\mathcal{O}}(\theta_1, \dots, \theta_n)_{\sigma_1, \dots, \sigma_n} \left(f_{\vartheta}^{\mathcal{O}'}(\theta_1, \dots, \theta_n)_{\sigma_1, \dots, \sigma_n}\right)^* \nonumber \\
  &\times  \exp\left({\rm i}x m\sum_{k=1}^n \sigma_k k(\theta_k) - {\rm i}tm \sum_{k=1}^n \sigma_k \varepsilon(\theta_k) \right),
\end{empheq}
\beq
\label{dressedtwopoint}
\eeq
where we define the filling fractions,
\beq
 \vartheta_{-1}(\theta_k)=\vartheta(\theta_k), \qquad \vartheta_{+1}(\theta_k)=\frac{\rho_h(\theta)}{\rho_p(\theta)}\vartheta(\theta_k).
\eeq
and
\begin{align}
  f_{\vartheta}^{\mathcal{O}}(\theta_1, \dots, \theta_n)_{\sigma_1, \dots, \sigma_n} = f_{\vartheta}^{\mathcal{O}}(\theta_1 + i\pi \delta_{\sigma_1,-1}, \dots, \theta_n + i\pi \delta_{\sigma_n,-1}).
\end{align}
The two-point function is then written as an expansion in terms of the dressed form factors, with particle and hole excitations on top of the background, $\vartheta$. 

The expression (\ref{dressedtwopoint}) can be contrasted with the LeClair-Mussardo conjecture for the finite-temperature two-point function:
\begin{align}
C_\vartheta^{\mathcal{O},\mathcal{O}^\prime}(x,t)=& \sum_{n=0} \sum_{\sigma_i = \pm 1}\left(\prod_{k=1}^n \int_{-\infty}^{\infty} \frac{d\theta}{2\pi}\vartheta_{\sigma_k}(\theta_k) \right) f^{\mathcal{O}}(\theta_1, \dots, \theta_n)_{\sigma_1, \dots, \sigma_n} \left(f^{\mathcal{O}'}(\theta_1, \dots, \theta_n)_{\sigma_1, \dots, \sigma_n} \right)^*\nonumber \\
  &\times  \exp\left(ix m\sum_{k=1}^n \sigma_k k(\theta_k) - itm \sum_{k=1}^n \sigma_k \epsilon(\theta_k) \right),
\end{align}
where only the zero-background form factors are used, and therefore the conjecture is not expected to be correct \cite{LECLAIR1999624}. Furthermore, the original LeClair-Mussardo proposal did not show how to properly regularize singularities arising from annihilation poles, while our expression (\ref{dressedtwopoint}) is fully regularized.

\section{Infrared regimes of the correlation function}\label{sec:infrared}

Form factor expansions of correlation functions are  known to be rapidly convergent in the infrared limit, in the zero-background case. In this case infrared refers to the limit of large separations between the  operators (compared to the natural length scale defined by the mass gap, $m$). Defining the Euclidean distance, $r=\sqrt{x^2+t^2}$, the zero-background two-point function may be written as
\begin{align}
  \langle 0 \vert \mathcal{O}(x,t)\mathcal{O}(0,0)\vert 0\rangle =& \sum_{n=0}^{\infty} \left(\prod_{k=1}^n \int_{-\infty}^{\infty} \frac{d\theta}{2\pi}\right) f^{\mathcal{O}}(\theta_1, \dots, \theta_n) f^{\mathcal{O}'}(\theta_1, \dots, \theta_n)  \nonumber \\
  &\times  \exp\left(-r m\sum_{k=1}^n\cosh\theta_k \right).
\end{align}
At large values of $mr$, it is easy to see that  this correlator decays exponentially as $e^{-mr}$, and the insertion of intermediate states with higher number of particles leads to corrections in powers of $e^{-mr}$. The exponential supression leads to this expansion being very quickly convergent.

The leading exponentially supressed contribution comes from the one-particle form factor, and can be written explicitly as
\beq
\langle 0 \vert \mathcal{O}(x,t)\mathcal{O}(0,0)\vert 0\rangle-\langle 0\vert\mathcal{O}(0,0)\vert 0\rangle^2=\frac{\vert f^\mathcal{O}_1\vert^2}{\pi}K_0(m r)+\mathcal{O}(e^{-2mr})\approx\frac{\vert f^\mathcal{O}_1\vert^2 }{\sqrt{2\pi mr}}e^{-mr},\nonumber
\eeq
for large $mr$, where $K_\alpha(x)$ is a modified Bessel function of the second kind.

We now consider the correlation function in the presence of a background density, $\vartheta(\theta)$, (\ref{dressedtwopoint}). In this case, there are two distinct IR limits we can consider. The first IR limit corresponds again to considering large separations between operators, or large values of $mr$. The second IR limit is the limit of small particle density, or small $\vartheta(\theta)$ for all $\theta$. 

Let us discuss first in some details what happens for large separtion between the operators. In such situations the factors
 \beq
 \exp\left({\rm i} x m\sum_{i=1}^n \sigma_i k(\theta_i)-{\rm i} t m\sum_{i=1}^n\sigma_i  \varepsilon(\theta_i)\right),\label{rsupress}
\eeq
 become highly oscillatory.  We then face different scenarios. First, there might be a saddle point of the exponent which then localizes contribution to the correlation function to the vicinity of the saddle point configuration. This happens when we consider asymptotes of the autocorrelation ($x= 0$, large $t$) or large distance and time asymptotics with fixed ratio $x/t$. This leading behaviour is then due to particle-hole configuration which minimizes the exponent.

However, there might be no saddle-point, as is the case for large $x$ asymptote of an equal-time correlation function. Then still the particle-hole contribution arises then as the leading one, but now this is caused by the form factor itself. The form-factors have a kinematic pole whenever $\theta_i \rightarrow \theta_j + i \pi$. The contribution from the pole is excluded by the regularization procedure. However, in the vicinity of the pole the derivative of the form-factor squared is not continous. Therefore, the integration picks up contribution whenever particle and hole are close to each other. The leading term is therefore again controlled by the particle-hole contributions.

There might another source of non-smoothness of the integrand: a discontinuity of the filling function $\vartheta(\theta)$. This is not the case at finite temperature, where the filling function is smooth, but e.g. at finite chemical potential. We don't consider this situation in the detail focusing solely on the finite temperature. In summary, for the thermal like states, the first infrared regime controlled by \eqref{rsupress} is dominated by the particle-hole contributions. 

Let us now consider the second infrared regime, when the filling function $\vartheta(\theta)$ is small. To be concrete we concentrate on one simple and physically interesting example: the thermal equilibrium, with a finite temperature, $T=1/\beta$. In this case, the function $\vartheta(\theta)$ can be derived by minimizing the entropy of the system, subject to a constraint fixing the total energy density. The particle occupation number function is then given by \cite{ZAMOLODCHIKOV1990695}
\beq
\vartheta(\theta)=\frac{1}{1+e^{\beta \varepsilon(\theta)}},\label{thermaloccupation}
\eeq
where $\varepsilon(\theta)$ is the dressed particle energy, which is determined as the solution of the integral equation
\beq
\beta\varepsilon(\theta)=\beta m\cosh\theta - \int \frac{d\theta^\prime}{2\pi} \varphi(\theta-\theta^\prime) \log\left(1+e^{-\beta\varepsilon(\theta^\prime)}\right).\nonumber
\eeq
At very low temperatures, or large $m\beta$, the function (\ref{thermaloccupation}) becomes small, and exponentially suppressed for all values of $\theta$. One can then express the low temperature two-point function as an expansion in powers of $e^{-m\beta}$. The suppresion factor arises then through
 \beq
 \prod_{i=1}^n\vartheta_{\sigma_i}(\theta_i),\label{betasupress}
 \eeq
in the integrand in~\eqref{dressedtwopoint}.

It is important to notice that the two different factors that lead to suppressions in the IR limits,  (\ref{betasupress}) and (\ref{rsupress}), have different behaviors, depending on whether the set of excitations are particles or holes, that is, depending on the signs $\sigma_i$. At large $m\beta$, the factor (\ref{betasupress}) leads to exponential supression of holes, while particle excitations are not supressed.  On the other the space-time asymptotes are controlled by the particle-hole pairs and states with only particles or holes are supressed by oscillating factors (\ref{rsupress}).

Determining what are the leading contributions in the IR limit of the expansion (\ref{dressedtwopoint}) depends on whether $m\beta$ or $mx$  is larger.  In the doubly IR limit, with $m\beta\to\infty$, and $mx\to\infty$, the two-point function (\ref{dressedtwopoint}) becomes simply
\beq
\lim_{\beta,x\to\infty}C_\vartheta^{\mathcal{O},\mathcal{O}}(x,t)=\left(\frac{\langle\vartheta\vert \mathcal{O}(0,0)\vert\vartheta\rangle}{\langle\vartheta\vert\vartheta\rangle}\right)^2,\label{zerothorder}
\eeq
that is, only the disconnected piece of the correlator remains. We can then examine what are the leading corrections to (\ref{zerothorder}) in the two IR regimes.

In the IR regime where $m\beta\gg mx$, the thermal corrections, $e^{-m\beta}$ are much smaller than the spatial corrections, $e^{-m x}$, therefore we can neglect contributions from hole excitations. The leading correction to (\ref{zerothorder}) is then given by the one-particle intermediate state,
\beq
C_\vartheta^{\mathcal{O},\mathcal{O}}(x,t)\approx\left(\frac{\langle\vartheta\vert \mathcal{O}(0,0)\vert\vartheta\rangle}{\langle\vartheta\vert\vartheta\rangle}\right)^2+\int_{-\infty}^\infty\frac{d\theta}{2\pi}\vartheta_+(\theta)\vert f_\vartheta^\mathcal{O}(\theta)\vert^2 e^{{\rm i}m x k(\theta)}.\label{regimeone}
\eeq

In the IR regime where $m\beta\ll mx$, contributions from states with only particles or only holes are supressed exponentially at long distances, while contributions from mixed particle-hole states are not. The leading correction to (\ref{zerothorder}) is then given by one particle-hole pair,
\beq
C_\vartheta^{\mathcal{O},\mathcal{O}}(x,t)&\approx&\left(\frac{\langle\vartheta\vert \mathcal{O}(0,0)\vert\vartheta\rangle}{\langle\vartheta\vert\vartheta\rangle}\right)^2\nonumber\\
&&\nonumber\\
&&+\int_{-\infty}^\infty \frac{d\theta_h\,d\theta_p}{4\pi^2}\vartheta_-(\theta_h)\vartheta_+(\theta_p)\vert f_\vartheta^{\mathcal{O}}(\theta_h+{\rm i}\pi,\theta_p)\vert^2e^{{\rm i}xm[k(\theta_p)-k(\theta_h)]}.\label{regimetwo}
\eeq 
Correlation functions in this regime, exhibit effectively gapless behavior, since there is no energy threshold needed to create a particle-hole pair. Depending on the properties of the background, $\vartheta(\theta)$, the correlator may exhibit polynomial or exponential decay at long-distances.

The regime, where particle-hole pairs are the dominant corrections, is analogous to the so-called hydrodynamic regime, that has been previously discussed in the non-relativistic Lieb Lininger model \cite{1742-5468-2015-2-P02019}. This is the only IR regime in the non-relativistic theory, since the number of particles is constant, so only particle-hole pairs can be created. The IR regime described in (\ref{regimeone}) is not present in the non-relativistic theory, since it involves the creation of new particles. A proper non-relativistic limit of the QFT correlator in a thermal background, is then given by the limit of very large mass, $m$, particularly such that the supression factor (\ref{rsupress}) is much larger than the supression factor (\ref{betasupress}).

\subsection{Comparison with previous low-temperature expansions}

We now compare the infrared limits of our two-point function to a previously known low-temperature expansion.  The first terms in a low temperature expansion of the thermal two point function, in an integrable QFT were rigorously computed in \cite{Pozsgay:2010cr}, by using a finite-volume regularization to take care of kinematical singularities, order by order.

 This approach consists of a straightforward expansion of the thermal correlator in terms of the finite-volume form factors,
\beq
C^{\mathcal{O},\mathcal{O}}_\beta (x) =\frac{{\rm Tr} \,e^{-\beta H}\mathcal{O}(x)\mathcal{O}(0)}{{\rm Tr } \,e^{-\beta H}}=\lim_{L\to\infty}\frac{ \sum_{N,M}C_{NM}}{\sum_N Z_N},
\eeq
with
\beq
C_{NM}=\int \{d\theta\}_N\{d\theta\}_M e^{-\beta E_N}\,_L\langle N\vert \mathcal{O}(x)\vert M\rangle_L\,_L\langle M\vert \mathcal{O}(0)\vert N\rangle_L,\,\,\,\,\,\,\,\,\,Z_N=\int\{d\theta\}_N e^{-\beta E_N}\,_L\langle N\vert N\rangle_L,
\eeq
where $\vert N\rangle_L$ and $\vert M\rangle_L$ are $N$- and $M$-particle states, respectively, at finite volume. 
The main idea is that the kinematical singularities in the form factors in the numerator, $C_{NM}$ can be cancelled exactly, order by order, with analogous singularities arising from an expansion of $(\sum_N Z_N)^{-1}$. The terms from the numerator and denominator can be combined into terms that contain no singularities,  and are properly regularized in the $L\to\infty$ limit, a procedure known as the ``linked cluster expansion",
\beq
C^{\mathcal{O},\mathcal{O}}_\beta (x) =\sum_{N,M}D_{NM},\label{linkedcluster}
\eeq
where the details of these regularized terms $D_{NM}$ can be found in the original reference \cite{Pozsgay:2010cr}, and will not be discussed here. We only point out that the term $D_{NM}$ is suppressed at finite temperature by a factor of $e^{-N\beta m\cosh(\theta)}$. This low-temperature expansion consists on computing the first few orders in $N$, while the label $M$ controls the exponential suppression at long distances, $x$.

At low temperatures, the leading terms are
\beq
C^{\mathcal{O},\mathcal{O}}_\beta (x) \approx\sum_{M}D_{0M}+\sum_{M} D_{1M}+\dots,
\eeq
where the first sum, $\sum_{M}D_{0M}$ is simply the vacuum two-point function, and $\sum_{M} D_{1M}$ is the two-point function averaged on a one-particle state, and so on.

The two-point function formula we have presented, Eq. (\ref{dressedtwopoint}) involves a complete reorganization of terms, compared to the expansion (\ref{linkedcluster}). For example, it is easy to see that even the zeroth order term in our expansion, (\ref{zerothorder}), involves a complete resummation of infinite terms of the expansion (\ref{linkedcluster}). To fully explore the differences between the two expansions, we can introduce the short-hand notation for Eq. (\ref{dressedtwopoint}),
\beq
C^{\mathcal{O},\mathcal{O}}_\beta (x)=\sum_{H,P}G_{HP}, \label{hpexpansion}
\eeq
where $G_{HP}$ is defined as the term in the expansion (\ref{dressedtwopoint}), containing $H$ holes and $P$ particles. Each term in our expansion (\ref{hpexpansion}) involves an infinite resummation of terms with all powers of $e^{-\beta m}$, so that (\ref{hpexpansion}) is not a low-temperature expansion, and the indices $H,P$ are not in direct correspondence with $N,M$ in (\ref{linkedcluster}). The reason for the discrepancy is that the expansion (\ref{hpexpansion}) is not centered around the vacuum state, but around the excited state, $\vert\vartheta\rangle$, so it already starts from an infinite resumation of finite-temperature terms, even at zeroth order

The way we can make a connection between the expansions (\ref{linkedcluster}) and (\ref{hpexpansion}) is by expanding each of the terms $G_{HP}$ in powers of $e^{-\beta m}$, as
\beq
G_{HP}=\sum_NG^N_{HP}
\eeq
where $G^N_{HP}$ is the term in the expansion proportional to $e^{-N \beta m}$. The correspondence between our expansion (\ref{hpexpansion}) and the linked cluster expansion (\ref{linkedcluster}) is then given by the relation
\beq
\sum_M D_{NM}=\sum_{H,P} G^{N}_{HP},
\eeq
which clearly indicates a complete reorganization of leading terms, between both expansions.

This being said, the main advantage of our expansion, is indeed the fact that it is centered around a highly excited state, while (\ref{linkedcluster}) is centered around the vacuum. This means that even just computing a handful of leading terms in our expansion can give nonperturbative insight that is inaccessible to the expansion (\ref{linkedcluster}) unless one is somehow able to resum an infinite number of terms. One simple explicit example of this is the clear separation of the two regimes discussed in Eqs. (\ref{regimeone}) and (\ref{regimetwo}). Additionally, it is completely trivial from our expansion to verify the clustering property, 
\beq
\lim_{x\to\infty} C^{\mathcal{O},\mathcal{O}}_\beta (x)=\langle\mathcal{O}\rangle_\beta\langle O\rangle_\beta,
\eeq
since this is just the zeroth order in our expansion. For Linked-Cluster and low-temperature expansions, the proof of this clustering property was shown in \cite{Pozsgay2018}, but it was a highly non-trivial task.

\section{Sinh-Gordon model}\label{sec:SinhGordon}

In this section we materialize the ideas introduced above in a concrete example of an Integrable QFT, namely the Sinh-Gordon (ShG) model. The action of the ShG model is
\begin{equation}
  \mathcal{S} = \int {\rm d}^2 x \left(\frac{1}{2}(\partial_{\mu}\phi(x))^2 - \frac{m^2}{g^2}\cosh g\phi(x) \right),
\end{equation}
where $g$ is the interaction parameter. The particle content of the ShG theory is especially simple. There is only one type of particle, and these particles are also  their own antiparticles: the field $\phi(x)$ is real. Moreover there are no bound states in the spectrum. Therefore the scattering matrix is just a single function, given by \cite{1979PhLB...87..389A}
\begin{equation} \label{scattering_matrix_ShG}
  S(\theta) = \frac{\tanh\frac{1}{2}(\theta - i \pi B/2)}{\tanh\frac{1}{2}(\theta + i \pi B/2)}, \qquad S(0) = -1.
\end{equation}
The renormalized coupling constant $B(g)$ is 
\begin{equation}
  B(g) = \frac{2g^2}{8\pi + g^2}.\label{bdef}
\end{equation}

To formulate the axioms we need to know functions $R_{\vartheta}(\theta_n\vert \theta_1, \dots, \theta_n)$. The builiding block for them is the back-flow function which can be computed once the filling function $\vartheta(\theta)$ is known. We formulate the finite volume theory with periodic boundary conditions. Therfore we will use the bosonic back-flow function $F_B(\tilde{\theta}|\theta)$. We will focus on the case of thermal equilibrium. In this case the filling function follows from the standard  Thermodynamic Bethe Ansatz (TBA) which we now briefly recall.

\begin{figure}[h]
  \center
  \begin{subfigure}{0.48\textwidth}
    \includegraphics[scale=0.5]{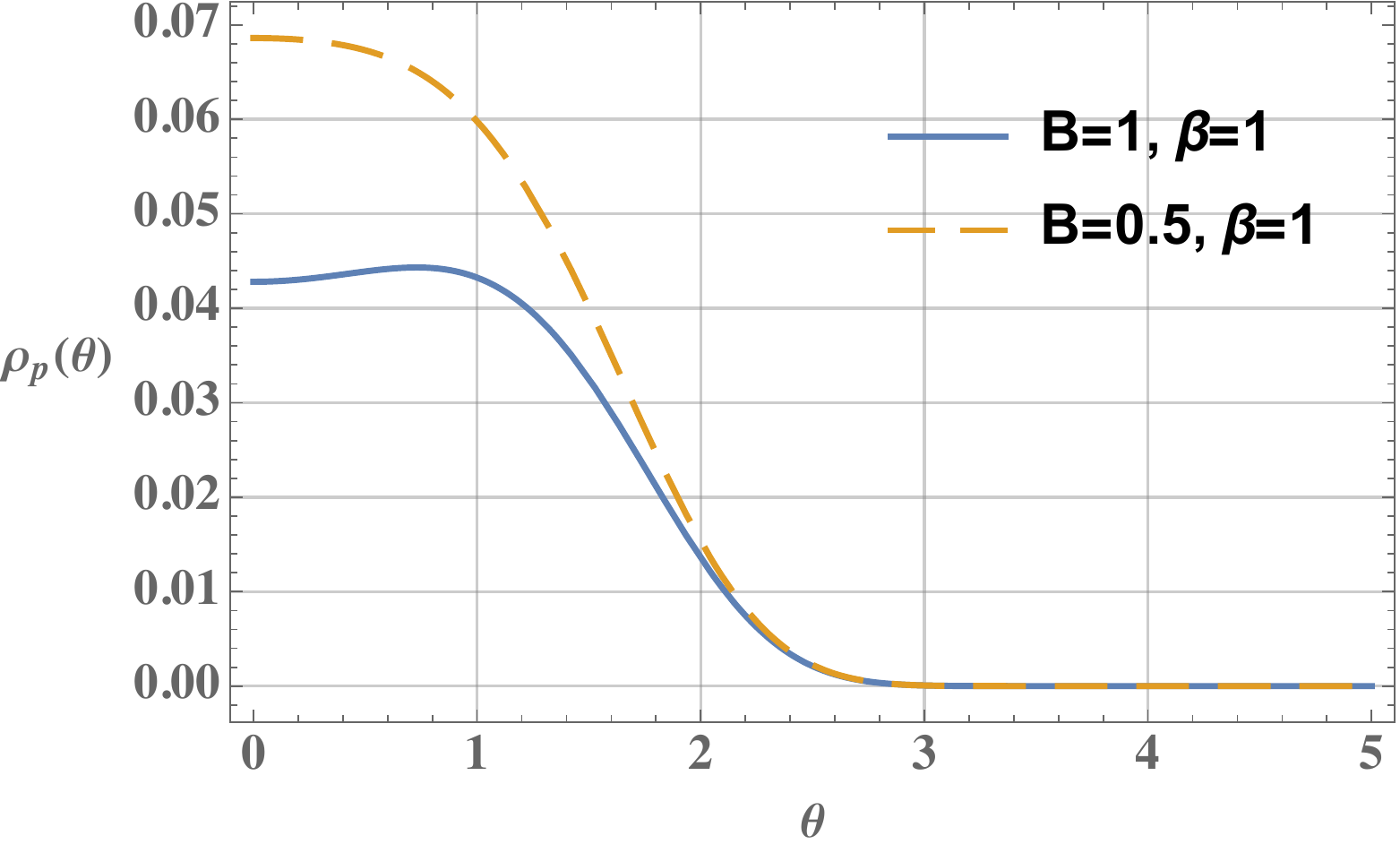}
  \end{subfigure}
  \begin{subfigure}{0.48\textwidth}
    \includegraphics[scale=0.5]{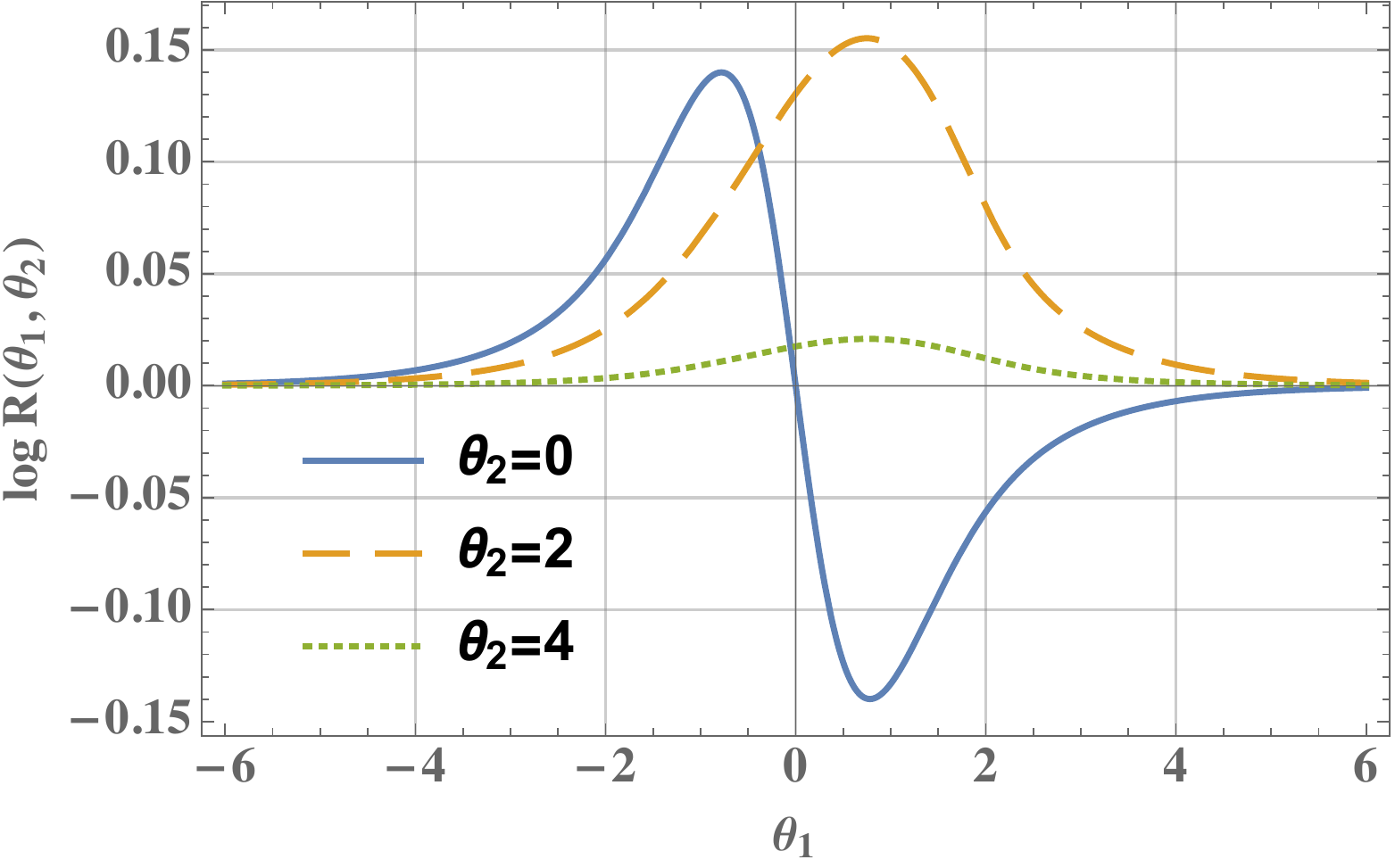}
  \end{subfigure}
  \caption{\emph{left panel:} Plots of the density $\rho_p(\theta)$ at different values of the interaction parameter and temperatures. \emph{right panel:} We plot the phase of the pure phase factor $R_{\vartheta}(\theta_1, \theta_2)$ for $B=0.5$ and $\beta=1$. }\label{fig:TBA}
\end{figure}

The thermal equilibrium at inverse temperature $\beta$ is described by the filling function
\begin{equation}
  \vartheta(\theta) = \frac{1}{1 + e^{\beta \varepsilon(\theta)}},
\end{equation}
with the dressed energy $\varepsilon(\theta)$ determined from the integral equation
\begin{equation}
  \beta\varepsilon(\theta)=\beta m\cosh\theta  - \int \frac{d\theta^\prime}{2\pi} \varphi(\theta-\theta^\prime) \log\left(1+e^{-\beta\varepsilon(\theta^\prime)}\right).
\end{equation}
Knowing the filling function, the particle density function $\rho_p(\theta)$ can be computed from eq.~\eqref{tba_rho}.
The crucial ingredient for the TBA is the kernel $\varphi(\theta)$ (the derivative of the phase shift). From eq.~\eqref{scattering_matrix_ShG} it follows that 
\begin{equation}
  \varphi(\theta) = \frac{\partial \delta(\theta)}{\partial \theta} = -i\frac{\partial}{\partial \theta}\log(-S(\theta)) = \frac{1}{\cos(\pi B/2 + i\theta)} + \frac{1}{\cos(\pi B/2 - i\theta)}.
\end{equation}
The TBA equations can be solved numerically. In Fig.~\ref{fig:TBA} we show the resulting particles density function $\rho_p(\theta)$ and the $R_{\vartheta}(\theta_1, \theta_2)$ factor.
Recall that the $R_{\vartheta}(\theta_1, \theta_2)$ factor is a pure phase function that describes the contribution of the background particles to the effective scattering between particles with rapidities $\theta_1$ and $\theta_2$. 

\begin{figure}
  \center
  \includegraphics[scale=0.65]{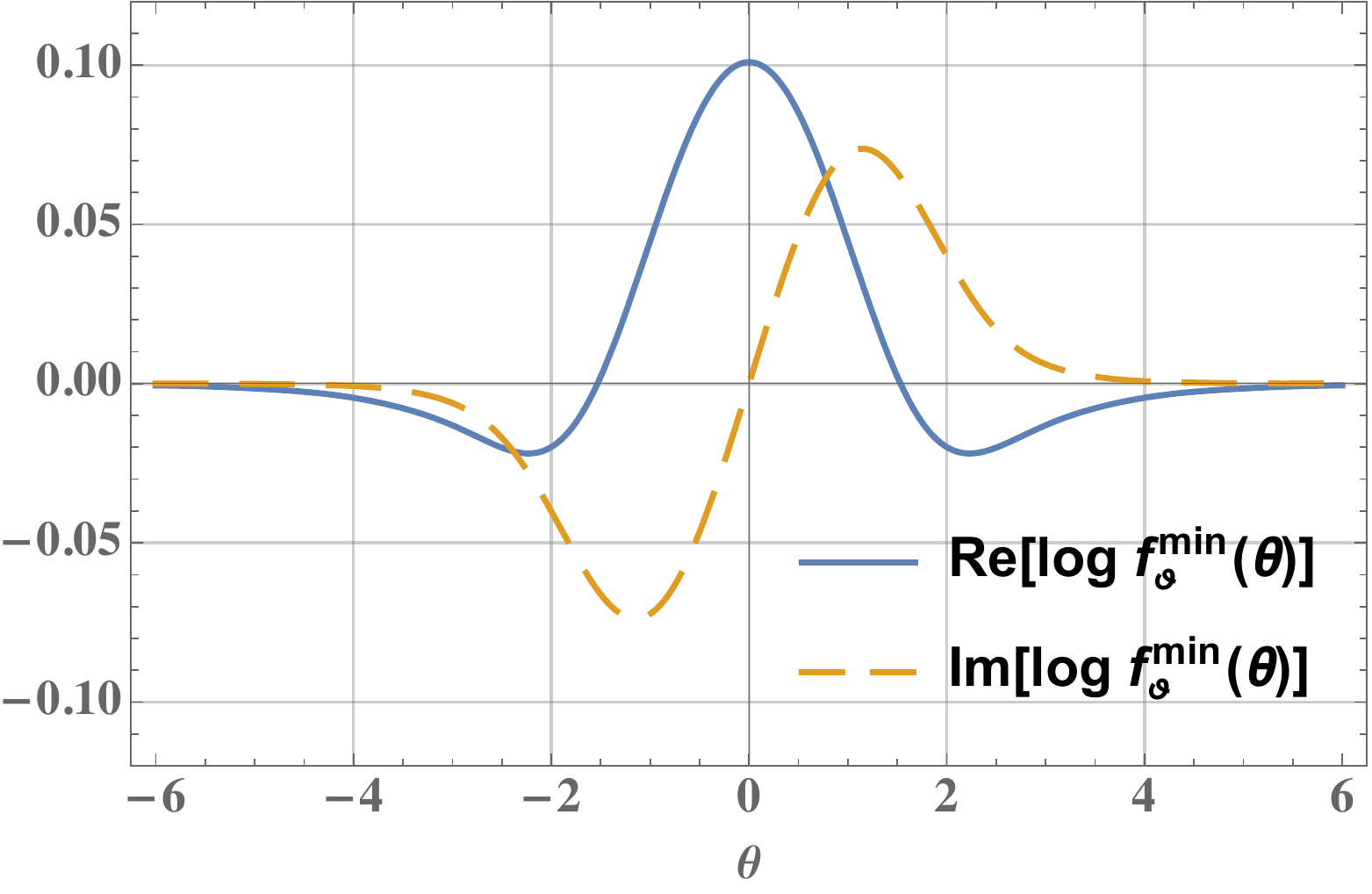}
  \caption{The one particle minimal form factor in the Sinh-Gordon theory for $B=0.5$ and $\beta = 1$. Contrary to the vacuum case, the one-particle form factor depends on the rapidity $\theta$. The asymptote $\lim_{|\theta|\rightarrow \infty} f_{\vartheta}^{\rm min}(\theta) = 1$. }\label{fig:onep_minimal}
\end{figure}

Equipped with the back-flow function we can now find the minimal solutions $f_{\vartheta}^{\rm min}(\theta_1, \dots, \theta_n)$ to the form-factor bootstrap. We follow the procedure outlined in section~\ref{sec:consequences}. We first analyze the one-particle form factors. We numerically find the representation~\eqref{definitionc} of $R_{\vartheta}(\theta|\theta)$ in terms of $C_{\vartheta}(t)$, that is
\begin{equation}
  C_{\vartheta}(t) = \frac{t}{\pi^2} \int d\theta \log R_{\vartheta}(\theta|\theta) \exp\left(\frac{i t\theta}{\pi} \right).
\end{equation}
Function $C_{\vartheta}(t)$ enters then the expression for the minimal form-factor
\begin{equation}
  f_\vartheta^{\rm min}(\theta)=\exp\left[-\frac{1}{4}\int_{-\infty}^\infty\frac{dt}{t}\frac{C_\vartheta(t)}{\sinh t}\exp\left(\frac{t(\theta - i\pi)}{i\pi}\right)\right].
\end{equation}
The results for the one-particle minimal form factor are presented in fig.~\ref{fig:onep_minimal}.

The general one-particle form factor takes the form
\begin{equation}
  f_{\vartheta}^{\mathcal{O}}(\theta) = K_{\vartheta}^{\mathcal{O}}(\theta) f_{\vartheta}^{\rm min}(\theta).
\end{equation}
Function $ K_{\vartheta}^{\mathcal{O}}(\theta)$ is a polynomial in $\cosh\theta$. The order of this polynomial can be bounded from above by considering the $\theta\rightarrow-\infty$ limit. The minimal form factor $f_{\vartheta}^{\rm min}(\theta)$ goes to $1$ in this limit. This implies that the polynomial must actually be a constant. This constant can still depend on the operator and on the state~$|\vartheta\rangle$. Therefore
\begin{equation}
  f_{\vartheta}^{\mathcal{O}}(\theta) = \mathcal{N}_{\vartheta}^{\mathcal{O}} f_{\vartheta}^{\rm min}(\theta).
\end{equation}
The constant can be found from the cluster properties. For an operator $\mathcal{O}_1$ that clusters into $\mathcal{O}_2\times\mathcal{O}_3$ the results is
\begin{equation}
  \mathcal{N}_{\vartheta}^{\mathcal{O}_1} = \mathcal{N}_{\rm cluster}^{\mathcal{O}_1}\,\frac{\langle\vartheta|\mathcal{O}_2|\vartheta\rangle}{\langle\vartheta|\vartheta\rangle} f_1^{\mathcal{O}_3}.
\end{equation}
Finally, the general form of background dependent one particle form factor in the Sinh-Gordon theory is
\begin{equation}
  f_{\vartheta}^{\mathcal{O}_1}(\theta) =\mathcal{N}_{\rm cluster}^{\mathcal{O}_1}\, \frac{\langle\vartheta|\mathcal{O}_2|\vartheta\rangle}{\langle\vartheta|\vartheta\rangle} f_1^{\mathcal{O}_3} f_{\vartheta}^{\rm min}(\theta),
\end{equation}
given that
\begin{equation}
  \mathcal{O}_1 \rightarrow \mathcal{O}_2 \times \mathcal{O}_3.
\end{equation}

We can now consider the particular case of vertex operators, $\mathcal{O}=\exp\left({\rm i}\alpha\phi\right)$, for some constant, $\alpha$. In this case it is particularly simple to fix the one-particle form factor for two reasons: vertex operators are self-clustering, and one-point functions of vertex operators are computable exactly through the so-called Negro-Smirnov formula \cite{Negro:2013wga}. 

The one-particle form factors of the vertex operator are then 
\beq
f_\vartheta^{e^{{\rm i}\alpha\phi}}(\theta)=\frac{1}{\langle 0\vert e^{{\rm i}\alpha\phi}\vert0\rangle}\frac{\langle \vartheta\vert e^{{\rm i}\alpha\phi}\vert\vartheta\rangle}{\langle\vartheta\vert\vartheta\rangle} f_1^{e^{{\rm i}\alpha\phi}}f_\vartheta^{\rm min}(\theta).\label{vertexonepoint}
\eeq
The three constants involved in expression (\ref{vertexonepoint}), namely $\langle 0\vert e^{{\rm i}\alpha\phi}\vert0\rangle$, $\frac{\langle \vartheta\vert e^{{\rm i}\alpha\phi}\vert\vartheta\rangle}{\langle\vartheta\vert\vartheta\rangle}$, and $f_1^{e^{{\rm i}\alpha\phi}}$ are exaclty known, and we include their expressions in Appendix \ref{appsinh}, for completeness. This completes the determination of one-particle form factors of vertex operators on a thermal background. We point out that it is easy to see that in the zero background limit, 
\beq
\lim_{\vartheta\to0}f_\vartheta^{e^{{\rm i}\alpha\phi}}(\theta)=f_1^{e^{{\rm i}\alpha\phi}},\nonumber
\eeq
as is expected.

Computation of form factors of two or more particles follows in a similar way. We first compute the function $\tilde{C}_{\vartheta}(t|\theta)$. It is given by
\begin{equation}
 \tilde{C}_{\vartheta}(t|\theta) = \frac{t}{\pi^2} \int_{-\infty}^{\infty} d\theta' \log R_{\vartheta}(\theta'|\theta) \exp\left(\frac{i t\theta'}{\pi} \right).
\end{equation}
The building block $A_{\vartheta}^{\rm min}(\theta;\theta_1, \dots, \theta_n)$ of the minimal form factor is then
\begin{equation}
  A_\vartheta^{\rm min}(\theta; \theta_1, \dots, \theta_n)=\exp\left[-\frac{1}{4}\int_{-\infty}^\infty\frac{dt}{t}\frac{C_\vartheta(t) + \tilde{C}_{\vartheta}(t|\theta_1, \dots, \theta_n)}{\sinh t}\exp\left(\frac{t(\theta - i\pi)}{i\pi}\right)\right].
\end{equation}
with $C_{\vartheta}(t|\theta_1, \dots, \theta_n) = \sum_{j=1}^n C_{\vartheta}(t|\theta_j)$. Finally, the expression for the minimal form factors comes from combining the minimal vacuum form factors with buliding block $A_{\vartheta}^{\rm min} (\theta; \theta_1, \dots, \theta_n)$. In the two particle case it is
\begin{equation}
  f_{\vartheta}^{\rm min}(\theta_1, \theta_2) = A_{\vartheta}^{\rm min}(\theta_1;\theta_2) A_{\vartheta}^{\rm min}(\theta_2, \theta_1) f^{\rm min}(\theta_1 - \theta_2).
\end{equation}
The standard vacuum minimal form factor $f^{\rm min}(\theta)$ in the Sinh-Gordon theory is
\begin{equation}
  f^{\rm min}(\theta) = \mathcal{N} \exp\left[8 \int_0^{\infty} \frac{dx}{x}\frac{\sinh\left(\frac{xB}{4}\right)\sinh\left(\frac{x}{2}\left(1 - \frac{B}{2}\right)\right)\sinh \frac{x}{2}}{\sinh x} \sin^2\left(\frac{x}{2\pi}(i\pi - \theta) \right)\right],
\end{equation}
with the normalization given by
\begin{equation}
  \mathcal{N} = \exp\left[-4\int_0^{\infty} \frac{dx}{x} \frac{\sinh\left(\frac{xB}{4}\right)\sinh\left(\frac{x}{2}\left(1 - \frac{B}{2}\right)\right)\sinh \frac{x}{2}}{\sinh^2 x} \right].
\end{equation}
This minimal form factor satisfies $\lim_{\theta\to\pm\infty} f^{\rm min}(\theta)=1$.
\begin{figure}
  \center
    \begin{subfigure}{0.45\textwidth}
    \includegraphics[scale=0.5]{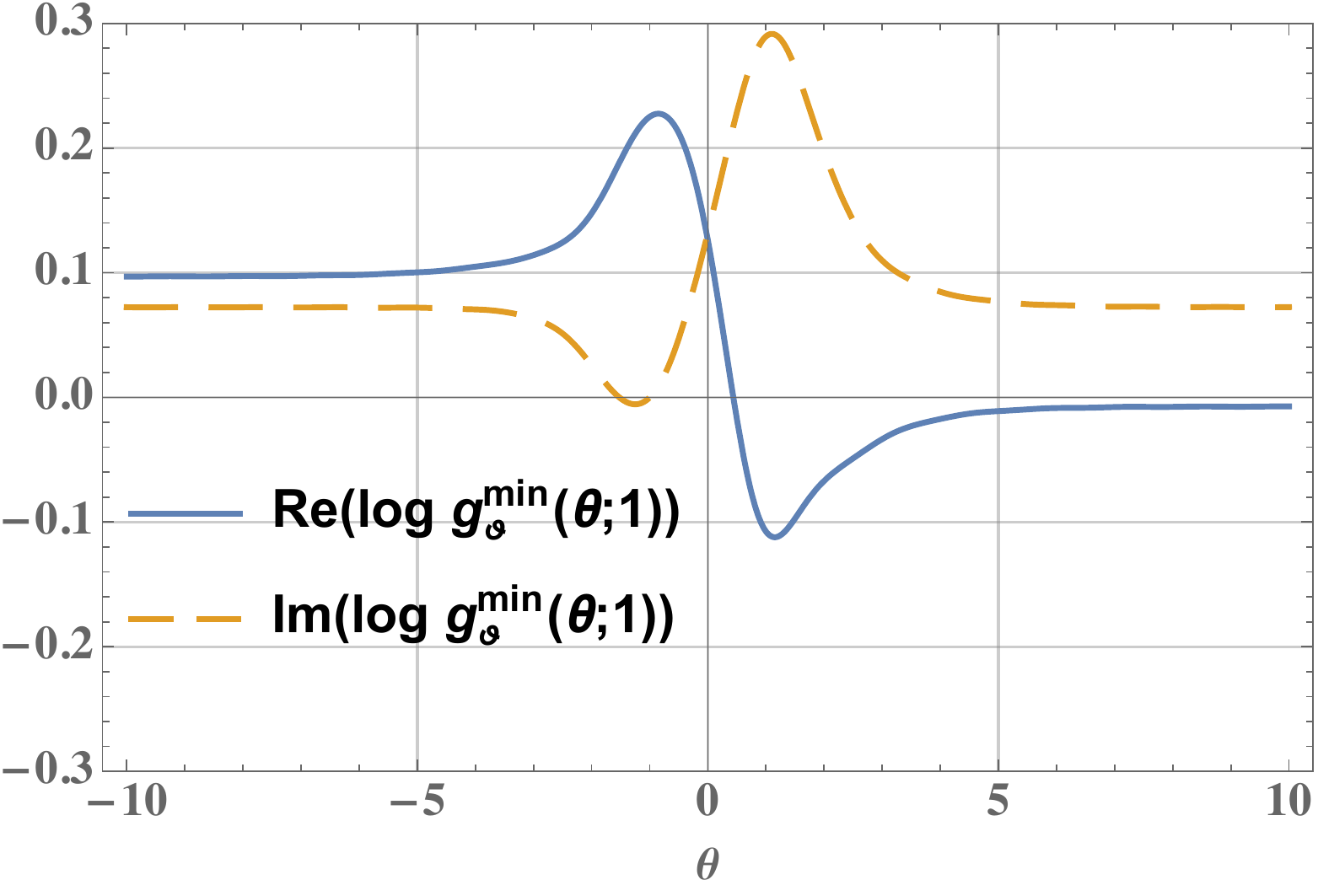}
  \end{subfigure}
    \begin{subfigure}{0.45\textwidth}
    \includegraphics[scale=0.5]{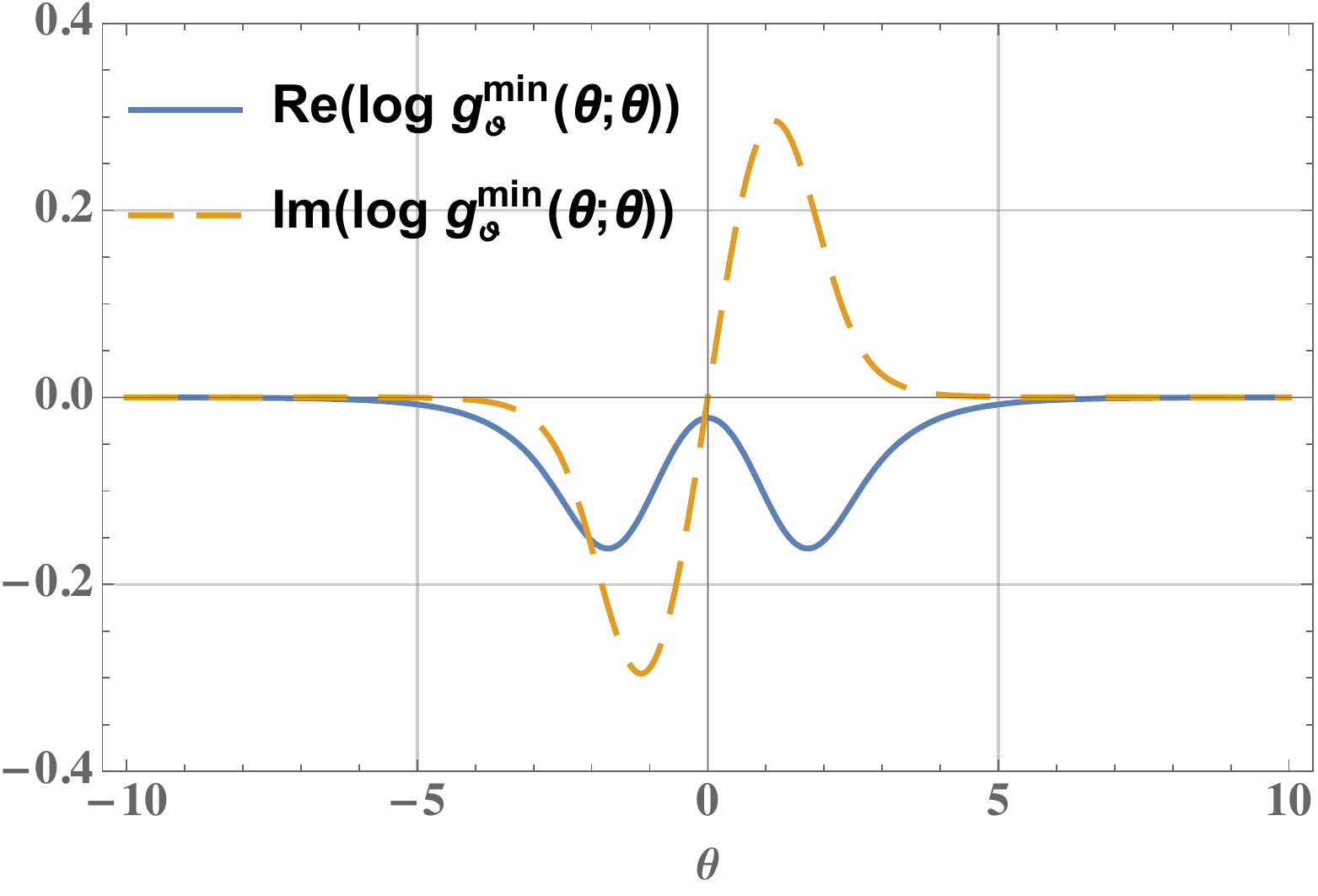}
  \end{subfigure}
  \caption{We plot the minimal two particle form factor normalized by the vacuum minimal form factor, defined as $g_\vartheta^{\rm min}(\theta_1,\theta_2)\equiv f_\vartheta^{\rm min}(\theta_1,\theta_2)/f^{\rm min}(\theta_1-\theta_2)$. Both plots are again for the Sinh-Gordon model with interaction parameter $B=1/2$ and at temperature~$\beta=1$.}\label{fig:twop_logs}
\end{figure}

Let us now consider the asymptotic of the background minimal form factor when $\theta_1 \rightarrow -\infty$. The problem boils down to determining the asymptotes of $A_{\vartheta}^{\rm min}(\theta_1, \theta_2)$  in both variables. We have
\begin{equation}
  \lim_{\theta_1\rightarrow - \infty} A_{\vartheta}^{\rm min}(\theta_1; \theta_2) = \exp\left[-\frac{\pi}{4} \frac{d}{dt}\tilde{C}_{\vartheta}(t|\theta_2)_{t=0} \right].
\end{equation}
The limit with the second variable approaching the inifinity is
\begin{equation}
  \lim_{\theta_1\rightarrow-\infty}A_{\vartheta}^{\rm min}(\theta_2, \theta_1) = A_{\vartheta}^{\rm min}(\theta_2).
\end{equation}
This shows that the asymptote of the minimal two particle form factor is
\begin{equation}
  \lim_{\theta_1\rightarrow-\infty}f_{\vartheta}^{\rm min}(\theta_1, \theta_2) = A_{\vartheta}^{\rm min}(\theta_2) \exp\left[-\frac{\pi}{4} \frac{d}{dt}\tilde{C}_{\vartheta}(t|\theta_2)_{t=0} \right]. \label{asymptotes_f2}
\end{equation}

\begin{figure}
  \center
    \begin{subfigure}{0.45\textwidth}
    \includegraphics[scale=0.5]{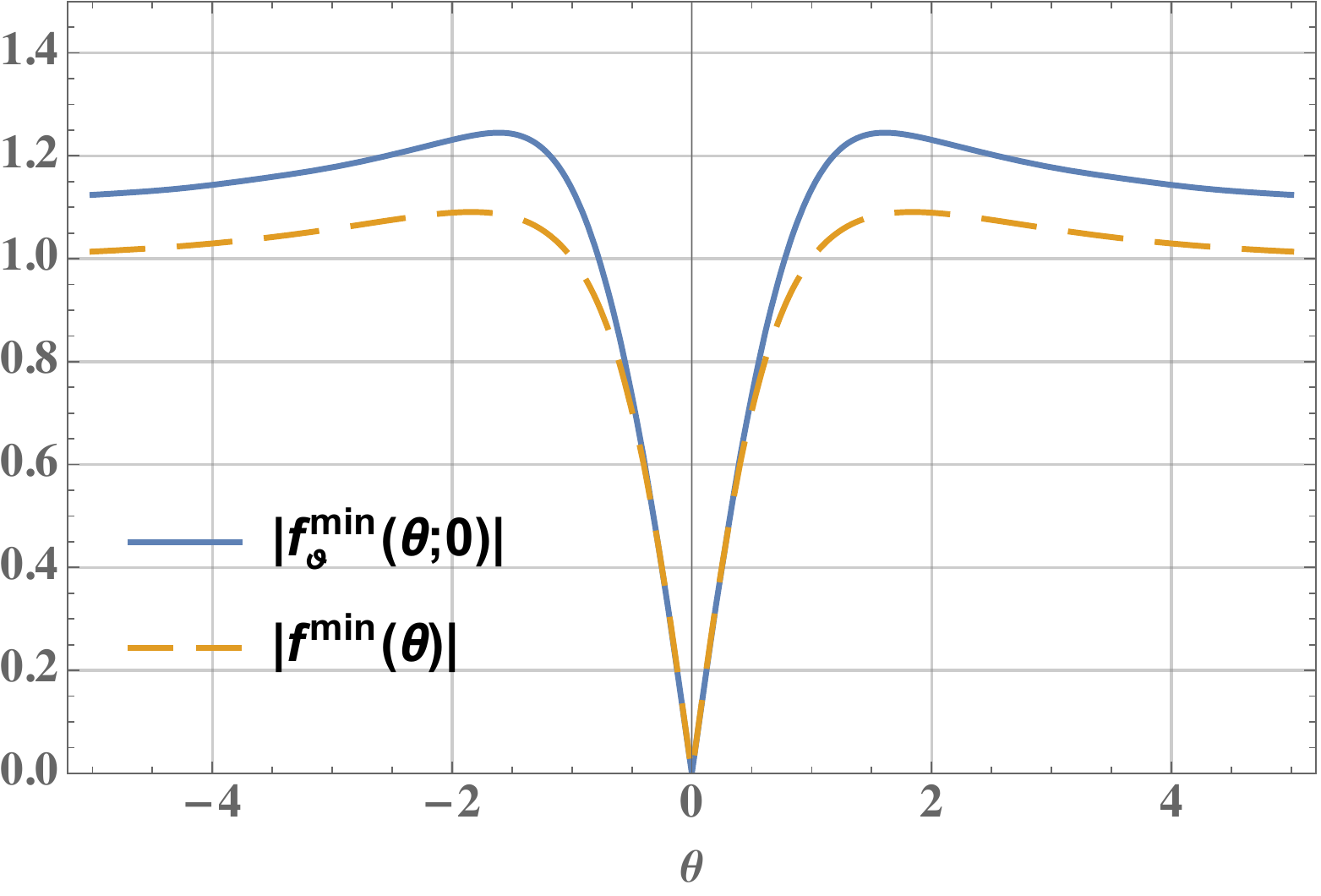}
  \end{subfigure}
    \begin{subfigure}{0.45\textwidth}
    \includegraphics[scale=0.5]{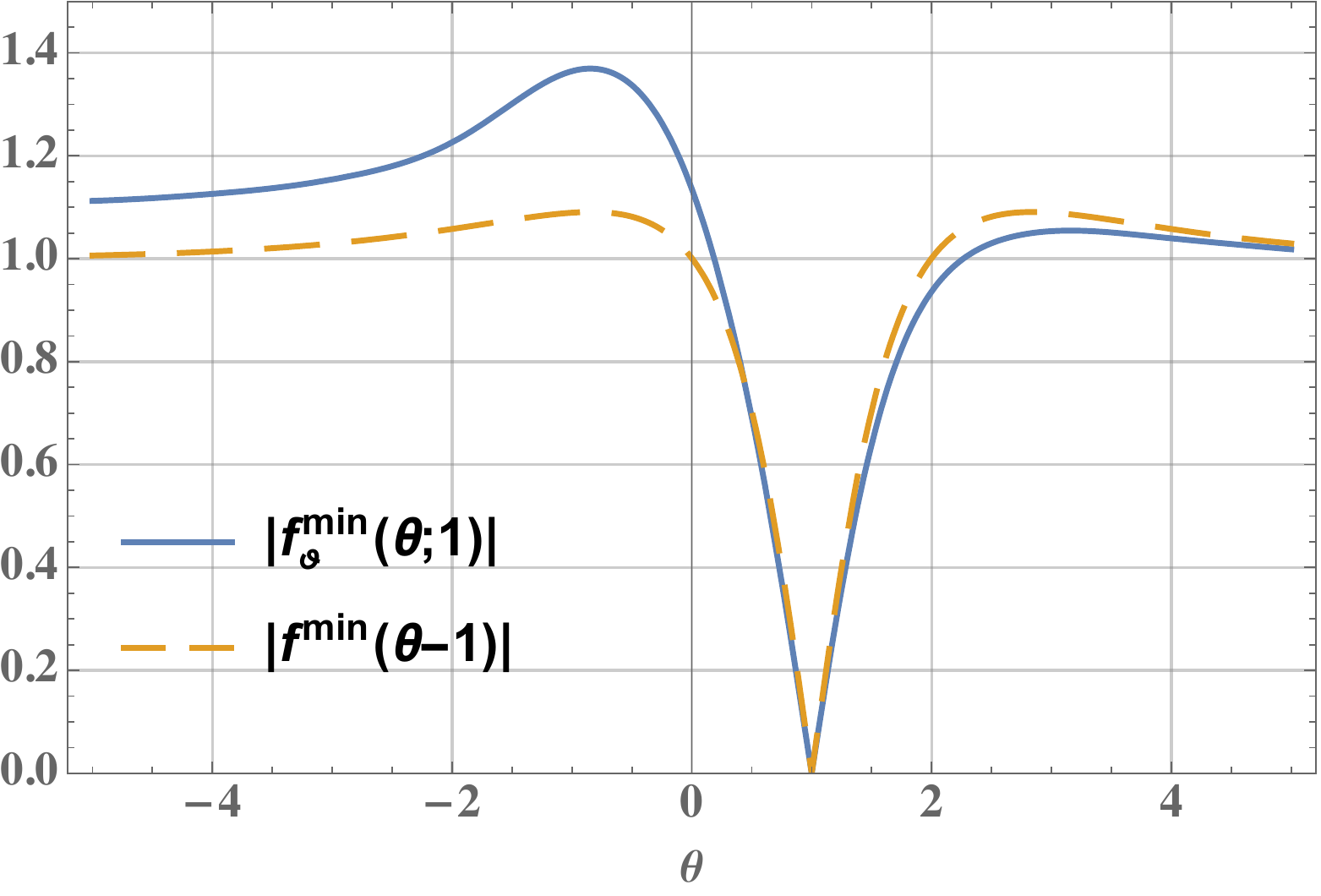}
  \end{subfigure}
  \caption{We plot the minimal two particle form factor over the background (of temperature $\beta =1$) and over the vacuum. Both plots are again for the Sinh-Gordon model with interaction parameter $B=1/2$.}\label{fig:twop_minimal}
\end{figure}

Figure~\ref{fig:twop_logs} shows the dressed two particle minimal form factor, $f^{\rm min}_\vartheta(\theta_1,\theta_2)$ normalized with respect to the vacuum minimal form factor $f^{\rm min}(\theta_1 - \theta_2)$, that is $f^{\rm min}_\vartheta(\theta_1,\theta_2)/f^{\rm min}(\theta_1 - \theta_2)$. We have checked that the asymptotes of $f^{\rm min}_\vartheta(\theta_1,\theta_2)$ agree with the analytical result~\eqref{asymptotes_f2}. Figure~\ref{fig:twop_minimal} shows the norm of the minimal form factor $|f^{\rm min}_\vartheta(\theta_1,\theta_2)|$ in comparison with the vacuum form factor $|f^{\rm min}(\theta_1-\theta_2)|$.

If we want to find form factors that satisfy the cluster properties, (\ref{cluster}), it is convenient then to re-normalize the minimal form factor as
\beq
\tilde{f}_\vartheta^{\rm min}(\theta_1,\theta_2)\equiv  \exp\left[\frac{\pi}{4} \frac{d}{dt}\tilde{C}_{\vartheta}(t|\theta_1)_{t=0} \right] \exp\left[\frac{\pi}{4} \frac{d}{dt}\tilde{C}_{\vartheta}(t|\theta_2)_{t=0} \right]f_{\vartheta}^{\rm min}(\theta_1,\theta_2),
\eeq
such that 
\beq
\lim_{\theta_2\to-\infty}\tilde{f}_\vartheta^{\rm min}(\theta_1,\theta_2)=f_\vartheta^{\rm min}(\theta_1).
\eeq
This renormalization is consistent with the form-factors axioms.

We consider now the vertex operators $e^{ {\rm i}\alpha \phi(x)}$. This form factor can be generally expressed as
\beq
f_\vartheta^{e^{{\rm i}\alpha\phi}}(\theta_1,\theta_2)=K_\vartheta^{e^{{\rm i}\alpha\phi }}(\theta_1,\theta_2) \tilde{f}_\vartheta^{\rm min}(\theta_1,\theta_2).
\eeq
The vertex operator is local, relative to the Sinh-Gordon particle, therefore there should be no annihilation pole in the function, $K_\vartheta^{e^{{\rm i}\alpha\phi }}(\theta_1,\theta_2) $. This function can then be expressed as a symmetric polynomial of the variables $x_1=e^{\theta_1},\,x_2=e^{\theta_2}$. 

The behavior of $f_\vartheta^{e^{{\rm i}\alpha\phi}}(\theta_1,\theta_2)$ is restricted at large rapidities by the bounds (\ref{boundongrowth}) and by the cluster property (\ref{cluster}). We have seen that this dressed minimal form factor goes to a constant value at large rapidites. This statement together with the bounds (\ref{cluster}), (\ref{boundongrowth}) imply that the symmetric polynomial can only be a constant, $K_\vartheta^{e^{{\rm i}\alpha\phi }}(\theta_1,\theta_2)\equiv K_{\vartheta\,\,12}^{e^{{\rm i}\alpha\phi }}$.

 The only task we have left is then to fix the constant $K_{\vartheta\,\,12}^{e^{{\rm i}\alpha\phi }}$. This can be done using the cluster property (\ref{cluster})
 \beq
 \lim_{a\to\infty} \tilde{f}_\vartheta^{e^{{\rm i}\alpha\phi}}(\theta_1-a,\theta_2-a)=\frac{1}{\langle 0\vert e^{{\rm i}\alpha\phi}\vert0\rangle}\frac{\langle \vartheta\vert e^{{\rm i}\alpha\phi}\vert\vartheta\rangle}{\langle\vartheta\vert\vartheta\rangle} f^{e^{{\rm i}\alpha\phi}}(\theta_1-\theta_2).\label{clustertwobreather}
 \eeq
 This expression (\ref{clustertwobreather}) can be further simplified by applying the cluster property again, to find 
 \beq
 \lim_{b\to\infty}f^{e^{{\rm i}\alpha\beta}}(\theta_1-\theta_2-b)=\frac{1}{\langle 0\vert e^{{\rm i}\alpha\phi}\vert0\rangle}\left(f^{e^{{\rm i}\alpha\phi}}_1\right)^2.
 \eeq
We can then fix the normalization constant as 
\beq
K_{\vartheta\,\,12}^{e^{{\rm i}\alpha\phi }}=\left(\frac{1}{\langle 0\vert e^{{\rm i}\alpha\phi}\vert0\rangle}\right)^2 \frac{\langle \vartheta\vert e^{{\rm i}\alpha\phi}\vert\vartheta\rangle}{\langle\vartheta\vert\vartheta\rangle}\left(f^{e^{{\rm i}\alpha\phi}}_1\right)^2,\label{solutiontwobreatherconstant}
\eeq
which completely fixes the dressed two-particle form factors of vertex operators.

\section{Conclusions and Outlook}

In this paper we have introduced a set of axioms for finite background form factors of local operators in Integrable Quantum Field Theory in 2d. These form factors are building blocks for the computation of correlation functions in states of finite energy density, such as finite temperature states. We have also found the minimal form factors which are the maximal analytic solutions to the axioms. These results open a path towards an explicit computation of the correlation functions in states of finite energy density.

Our work includes a number of simplifying assumptions. We focused on  integrable QFT's with diagonal scattering, with a single species of particles and without bound states. The last two assumptions are easier to generalize, since it should not be too difficult to introduce an additional bound-state bootstrap axiom in the presence of a thermodynamic background, which takes care of new bound-state poles appearing in the S-matrix. The generalization to non-diagonal scattering theories seems much more challenging at present, since the thermodynamic Bethe ansatz formulation of non-diagonal theories is much more complicated, and requires the introduction of additional magnonic and string-like pseudoparticles \cite{Bajnok:2010ke}. While non-diagonal theories are too difficult for this introductory work, it is definitely an interesting direction to study in the future, since some physically relevant QFT's, such as the sine-Gordon model, and the $O(N)$ sigma model, have non-diagonal scattering \cite{ZAMOLODCHIKOV1979253}. Developing further in this direction will allow us to compare our approach to the numerical results available from  TCSA computations \cite{Kukuljan:2018whw}.

Also for the sake of simplicity, we have chosen to study only operators which are local, relative to the particle excitations. This excludes theories which have soliton, or kink-like particles. While it is easy to generalize the standard form factor bootstrap on the vacuum to include semi-local operators, the generalization in the presence of a thermodynamical background is much more involved. This difficulty can be seen even in the simple case of Ising field theory \cite{Doyon:2005jf}, which is a theory of free Majorana fermions. The fermionic particles are not local relative to the spin operator, which makes the computation of finite-temperature form factors highly non-trivial, even for this non-interacting case. 

As a simplifying step in our calculation, in equation (\ref{nosoftmodes}) we assumed that the form factors on top of the background can be defined while ignoring the averaging over soft modes that lead to the same particle distribution. At very high energy densities, however, these soft modes may have a non-negligible effect, which can be interpreted as a renormalization of the form factor. At high energy densities, many soft particle-hole pairs (with momenta close to each other) can be created with no energy gap. Therefore the physical renormalized form factors would become those where one sums over soft particle-hole pairs leading to the same physical state. The computation of such form factor renormalization factors, which will be relevant at high temperatures, will be left for a future.

Formulating the axioms is of course only the first step. The next step is to actually determine form factors of interesting operators in a variety of theories. Here, we have shown, as a proof of principle, how to determine the one- and two-particle  form factors of vertex operators in the Sinh-Gordon model.  There exist many other integrable QFT's  of physical and mathematical interest, which have been, over the course of many years, studied through the standard form factor bootstrap program (there are too many examples to refer individually to all here, but for an introductory overview of the applications of the bootstrap program in a variety of field theories, see Ref. \cite{mussardobook}). After establishing our new set of axioms in the presence of a thermodynamic background, a large amount of work is set for the future, to re-derive the catalog of known exact form factors for the large number of IQFT's, but now in the presence of a thermodynamic background. 

The main concrete applications we discussed here concern the simplest kind of finite-energy-density background, namely, the case of thermal equilibrium. An important task for the future will also be to apply the program developed here to other interesting cases, such as computing correlation functions after a quantum quench, perhaps by taking advantage of the quench action approach \cite{PhysRevLett.110.257203}, which formulates non-equilibrium correlators in terms of representative eigenstates like the ones we considered here. 

Another physical problem to address in the future is that of inhomogeneous quantum quenches, and in particular the problem of quantum transport. It has been recently understood that properties of quantities such as the Drude weight \cite{2013CMaPh.318..809I}, which can be computed in terms of correlation fucntions in the presence of a spatially inhomogeneous background, can determine qualitative aspects of quantum transport in integrable systems, in particular, whether energy is transported ballistically or diffusively \cite{DeNardis:2018omc}. It would be interesting to study in the future these kind of correlations in the presence of an inhomogenous background, and give some insight into the transport properties of IQFT's.

Another interesting direction to consider in the future is the non-relativistic limit of the results derived in this paper. In particular, it is known that the Lieb-Liniger model can be obtained as a non-relativistic limit of the sinh-Gordon QFT, and several useful results have been recently derived using this correspondence \cite{Kormos:2009yp, 1742-5468-2016-12-123104,Bastianello:2018bbe}. It would be interesting to consider the non-relativistic limit of form factors derived for the sinh-Gordon model in the presence of a background, and see how (and if) we can recover thermodynamic form factors, for example, those that have been previously computed for the Lieb-Liniger model \cite{1742-5468-2015-2-P02019}.

We will conclude by remarking that the standard form factor bootstrap on a vacuum state has seen too many applications in a variety of fields to provide a complete list here. The program we have outlined in this paper, which generalizes the bootstrap program to arbitrary highly excited states, should then also be of a similarly wide applicability, as we can now begin to explore the same kind of applications of QFT correlation functions, but in a variety of physically relevant thermodynamic states.

\section*{Acknowledgments}

We thank Zoltan B{\'a}jnok, Jean-S{\'e}bastien Caux, M{\'a}rton Kormos, Jacopo de Nardis, Bal\'{a}zs Pozsgay, Tom Price, and Dirk Schuricht for stimulating discussions on the subject and comments on the manuscript.
ACC acknowledges support from  the European Union's Horizon 2020 under the Marie Sklodowoska-Curie grant agreement 750092, while MP is supported by the NCN under FUGA grant 2015/16/S/ST2/00448.

\newpage
\appendix

\section{Finite Volume Field Theory} \label{app:finite_volume}

In this appendix we derive  the integral equations for the densities and the back-flow functions. These results are not new. They are included for the convenience to the reader.

\subsection*{Density equation}

To derive the equation for the densities it is convenient to start with a finite size version of the density defined in terms of the rapidities solving the Bethe equations~\eqref{bethequantization}. 
\begin{equation}
  \rho_{p}(\theta) = \frac{1}{L}\sum_{l=1}^n \delta_D(\theta - \tilde{\theta}_l).
\end{equation}
Here $\delta_D$ is the Dirac's delta function. We also define a function $x(\theta)$ through
\begin{equation}
  2\pi x(\theta) = m \sinh \theta + \frac{1}{L}\sum_{l}\delta(\theta - \tilde{\theta}_l).
\end{equation}
From the definition it follows that $x(\tilde{\theta}_k) = I_j/L$. Computing $dx/d\theta$ we find
\begin{equation}
  2\pi \frac{dx}{d\theta} = m \cosh \theta + \frac{1}{L}\sum_{l\neq k}\varphi(\theta - \tilde{\theta}_l).
\end{equation}
On the other hand $\frac{dx}{d\theta}$ is the Jacobian of a mapping from the $x$-space to the $\theta$ space. Therefore it expresses the density of accessible rapidities in the vicinity of $\theta$, that is the $\rho_s(\theta)$. Taking the thermodynamic limit of the sum we find
\begin{equation}
  2\pi \rho_s(\theta) =  m \cosh \theta + \int d\theta' \rho_p(\theta') \varphi(\theta - \theta').
\end{equation}

\subsection*{Back-flow}

The back-flow function describes a shift in rapidities due to an introduction of an extra particle or a hole. The equation for the back-flow can be derived by comparing the Bethe equations~\eqref{bethequantization} for the background state $|\vartheta\rangle$ and an excited state $|\vartheta, \theta_1\rangle$. Let $\{\tilde{\theta}\}$ be a set of rapidities of a finite version of the $|\vartheta\rangle$ state and $\{\tilde{\theta}'\}$ together with $\theta$ be a set of rapidities of a finite version of $|\vartheta, \theta\rangle$. Compare now the Bethe equations for the corresponding rapidities in the two states. Because the quantum numbers are the same we get
\begin{equation}
  0 = mL \sinh \tilde{\theta}_k' - mL \sinh \tilde{\theta}_k + \sum_{l\neq k} \left(\delta_0(\tilde{\theta}_k' - \tilde{\theta}_l') - \delta_0(\tilde{\theta}_k - \tilde{\theta}_l)\right) + \delta(\tilde{\theta}_k' - \theta_1).
\end{equation}
For the equality to be fulfilled the difference $\tilde{\theta}_k' - \tilde{\theta}_k$ must be of order $1/L$. We parametrize it, anticipating the result, in the following way,
\begin{equation}
  \tilde{\theta}_k' - \tilde{\theta}_k = - \frac{F_F(\tilde{\theta}_k| \theta_1)}{L \rho_s(\tilde{\theta}_k)}.
\end{equation}
Substituting and expanding in inverse powers of $L$, in the leading order, we find
\begin{equation}
  0 = - \frac{F_F(\tilde{\theta}_k|\theta_1)}{\rho_s(\tilde{\theta}_k)}\left( m \cosh\tilde{\theta}_k + \frac{1}{L}\sum_{l \neq k} \varphi(\tilde{\theta}_k - \tilde{\theta}_l)\right) + \frac{1}{L}\sum_{l \neq k} \varphi(\tilde{\theta}_k - \tilde{\theta}_l) \frac{F(\tilde{\theta}_l|\theta_1)}{\rho_s(\tilde{\theta}_l)} + \delta_0(\tilde{\theta}_k - \theta_1).
\end{equation}
In the thermodynamic limit, the bracket becomes equal $2\pi \rho_s(\tilde{\theta}_k)$ and the second sum turns into an integral. Reorganizing we find
\begin{equation}
  2\pi F_F(\theta|\theta_1) = \delta_F(\theta - \theta_1) + \int d\theta' \vartheta(\theta')\varphi(\theta - \theta') F_F(\theta'|\theta_1), \qquad \delta_F(\theta) = \delta_0(\theta).
\end{equation}
This is the equation for the fermionic back-flow function. The bosonic back-flow function $F_B(\theta|\theta_1)$ can be found in analogous way.

The back-flow can be used to express the energy $\epsilon(\theta)$ and momentum $k(\theta)$ of the excited state $|\vartheta, \theta\rangle$ with respect to the background state $|\vartheta\rangle$. From the finite-size expressions for the energy and momentum, in the thermodynamic limit, we find
\begin{align}
  k(\theta) &= m\sinh\theta - \int d\theta' \vartheta(\theta') \cosh \theta' F(\theta'|\theta),\\
  \epsilon(\theta) &= m\cosh\theta -  \int d\theta' \vartheta(\theta') \sinh \theta' F(\theta'|\theta).
\end{align}
The energy and momentum of a state with more excitations is the sum of the energy and momentum carried by each single excitation. At the thermal equilibrium and for fermions, these functions can be expressed in a standard TBA~\cite{ZAMOLODCHIKOV1990695} form
\beq
\varepsilon(\theta)&=&m\cosh\theta+\int \frac{d\theta^\prime}{2\pi}\varphi(\theta-\theta^\prime)\vartheta(\theta^\prime)\varepsilon(\theta^\prime),\nonumber\\
k(\theta)&=&m\sinh\theta+\int \frac{d\theta^\prime}{2\pi}\varphi(\theta-\theta^\prime)\vartheta(\theta^\prime)k(\theta^\prime).
\eeq

\section{Normalization constants in Sinh Gordon}\label{appsinh}

In this appendix, for the sake of completeness, we list several previously derived formulas that we make use of, to normalize the one- and two-particle form factors of vertex operators in the Sinh-Gordon model.

In both, the one- and two-particle cases, the normalization constants were found in terms of three values, namely, the vacuum expectation value of the vertex operator, the one-particle form factor (on top of a vacuum state), and the one-point function at finite background, of the same operator.

An integral expression for the vacuum expectation value was found in \cite{Lukyanov:1996jj}, and it is given by
\beq
\langle 0\vert e^{{\rm i}\alpha\phi}\vert0\rangle=\left[\frac{m\Gamma\left(\frac{1}{gq}\right)\Gamma\left(1+\frac{g}{2q}\right)}{4\sqrt{\pi}}\right]^{2\alpha^2}\exp\left\{\int_0^\infty\frac{dt}{t}\left[-\frac{\sin^2(2\alpha g t)}{2\sinh(g^2t)\sinh(t)\cosh(qgt)}-2\alpha^2e^{-2t}\right]\right\}
\eeq
where $q=g+g^{-1}$.

An integral expression has also been found for the one-particle form factor in \cite{Lukyanov:1997bp}, and it is given by
\beq
f_1^{e^{{\rm i}\alpha\phi}}=2\,\langle 0\vert e^{{\rm i}\alpha\phi}\vert0\rangle\,\cos\frac{\pi\xi}{2}\sqrt{2\sin\frac{\pi\xi}{2}}\,\frac{\sin\left(\frac{\pi\xi\alpha}{g}\right)}{\sin(\pi\xi)}\exp\left\{\int_0^{\pi\xi}\frac{dt}{2\pi}\frac{t}{\sin(t)}\right\}.
\eeq
where $\xi=g^2/(1+g^2)$.

The computation of the expectation value in the presence of the background can be performed in two ways. The first approach is through the one-point Leclair-Mussardo formula \cite{LECLAIR1999624}, which has the advantage of being easily generalizable to any local operator of any field theory with diagonal scattering. This Leclair-Mussardo formula, for a given operator, $\mathcal{O}(x)$, and a background  given by the distribution, $\vartheta(\theta)$, is given by 
\beq
\frac{\langle \vartheta\vert\mathcal{O}\vert \vartheta\rangle}{\langle\vartheta\vert\vartheta\rangle}=\sum_{n=0}^\infty \frac{1}{n!}\int_{-\infty}^\infty\dots\int_{-\infty}^\infty\prod_{i=1}^n\left[\frac{d\theta_i}{2\pi}\vartheta(\theta_i)\right]\langle \theta_n,\dots,\theta_1\vert\mathcal{O}(0)\vert\theta_1,\dots,\theta_n\rangle_c,\nonumber
\eeq
where the subscript '$c$' in the matrix elements denotes that we are considering only the connected part of the form factors, defined as
\beq
\langle \theta_n,\dots,\theta_1\vert\mathcal{O}(0)\vert\theta_1,\dots,\theta_n\rangle_c=\lim_{\{\zeta_i\}\to0}\langle 0\vert \mathcal{O}\vert \theta_1,\dots,\theta_n,\theta_n-{\rm i}\pi+{\rm i}\zeta_n,\dots,\theta_1-{\rm i}\pi+{\rm i}\zeta_1\rangle\vert_{\rm finite\,part}.
\eeq
The disadvantage of the Leclair-Mussardo formula is that, in practice, it can only be evaluated as a low-energy density expansion, by computing one by one the few-particle form factor terms. It is not clear if a complete resummation of all terms is possible, to yield any closed form expression valid at higher energy densities.

A more convenient expression is available, proposed by Negro and Smirnov \cite{NEGRO2013166,Negro:2014nva}, exclusively for the finite-background expectation values of vertex operators in the sinh-Gordon model. While being specific to one kind of operator in a particular QFT, the Negro-Smirnov formula has the advantage of being a fully resummed integral expression, that does not need to be computed order by order like the Leclair-Mussardo formula. The one-point functions are given by the integral expression
\beq
\frac{\langle \vartheta\vert e^{(k+1)g\phi}\vert\vartheta\rangle}{\langle \vartheta \vert e^{kg\phi}\vert \vartheta\rangle}=1+\frac{2\sin(\pi B(k+1/2))}{\pi}\int_{-\infty}^\infty d\theta\, \vartheta(\theta)\,e^{\theta}p_k(\theta),\label{negrosmirnov}
\eeq
where
\beq
p_k(\theta)=e^{-\theta}+\int_{-\infty}^\infty d\mu \,\vartheta(\mu) \,\chi_k(\theta-\mu)p_k(\mu),\nonumber
\eeq
and
\beq
\chi_k(\theta)=\frac{{\rm i}}{2\pi}\left(\frac{e^{-{\rm i} kB\pi}}{\sinh(\theta+{\rm i}\pi\alpha)}-\frac{e^{{\rm i}k B\pi}}{\sinh(\theta-{\rm i}\pi \alpha)}\right).\nonumber
\eeq
and $B(g)$ is defined in Eq. (\ref{bdef}).

The formula (\ref{negrosmirnov}) only computes ratios of expectation values of vertex operators. Nevertheless, as was pointed out in \cite{Bertini:2016xgd}, if $B(g)$ is an irrational number, this formula can be used to compute the expectation value of the vertex operator for arbitrary $k$, in a procedure outlined in \cite{Bertini:2016xgd}.

\bibliographystyle{JHEP}
\bibliography{biblio}

\end{document}